\documentclass{aa}  
\usepackage{graphicx}
\usepackage{txfonts}
\begin{document}
\title{Kinetic Simulation of the Electron-Cyclotron Maser Instability: Relaxation of Electron Horseshoe Distributions}
\titlerunning{Simulation of the Electron-Cyclotron Maser Instability}
\author{A.A. Kuznetsov}
\authorrunning{A.A. Kuznetsov}
\institute{Armagh Observatory, Armagh BT61 9DG, Northern Ireland\\
           \email{aku@arm.ac.uk}
           \and
           Institute of Solar-Terrestrial Physics, Irkutsk 664033, Russia}
\date{Received *; accepted *}
\abstract{The electron-cyclotron maser instability is responsible for generation of the auroral kilometric radiation of the Earth and similar phenomena at other magnetized planets of the Solar system. Recently discovered radio emission from ultracool dwarfs has many similarities with the planetary auroral radio emissions. The in situ measurements in the terrestrial magnetosphere indicate that the radiation results from nonthermal electrons with a horseshoe-like distribution. Kinetic simulations of the electron-cyclotron maser instability for such distribitions have not been done yet.}%
{In this work, we investigate amplification of plasma waves by the horseshoe-like electron distribution as well as relaxation of this distribution due to the electron-cyclotron maser instability. We aim to determine the parameters of the generated plasma waves, timescales of the relaxation process, and the conversion efficiency of the particle energy into waves.}%
{We have developed a kinetic relativistic quasi-linear 2D code for simulating the coevolution of an electron distribution and the high-frequency plasma waves. The code includes the processes of wave growth and particle diffusion which are assumed to be much faster than other processes (particle injection, etc.). A number of simulations have been performed for different parameter sets which seem to be typical for the magnetospheres of ultracool dwarfs (in particular, the plasma frequency is much less than the cyclotron one).}%
{The calculations have shown that the fundamental extraordinary mode dominates strongly. The generated waves have the frequency slightly below the electron cyclotron frequency and propagate across the magnetic field. The final intensities of other modes are negligible. The conversion efficiency of the electron energy into the extraordinary waves is typically around 10\%. Complete relaxation of the unstable electron distribution takes much less than a second.}%
{Energy efficiency of the electron-cyclotron maser instability is more than sufficient to provide the observed intensity of radio emission from ultracool dwarfs. On the other hand, the observed light curves of the emission are not related to the properties of this instability and reflect, most likely, dynamics of the electron acceleration process and/or geometry of the radiation source.} 
\maketitle

\section{Introduction}
The electron-cyclotron maser instability (ECMI) can occur in a magnetized plasma with a non-equilibrium electron distribution function (a positive slope in perpendicular velocity is required) and results in an effective amplification of electromagnetic waves at the low harmonics of the electron cyclotron frequency (Wu \& Lee \cite{wu79}; Melrose \& Dulk \cite{mel82}; Winglee \& Dulk \cite{win86}). In a relatively rarefied plasma (when the electron plasma frequency to cyclotron frequency ratio  $\omega_{\mathrm{p}}/\omega_{\mathrm{B}}\ll 1$), the dominant mode of astrophysical electron-cyclotron masers is the extraordinary wave with the frequency close to the fundamental cyclotron frequency. The emission has a narrow spectral bandwidth, a narrow directivity pattern, and high (nearly 100\%) degree of circular polarization. ECMI is responsible for generation of the auroral kilometric radiation (AKR) in the magnetosphere of the Earth and, most likely, the similar phenomena at other magnetized planets of the Solar system (Zarka \cite{zar98}; Treumann \cite{tre06}). ECMI has also been applied to interpret certain types of solar and stellar sporadic radio bursts (Melrose \& Dulk \cite{mel82}; Fleishman \& Melnikov \cite{fle98}).

Recently, a number of late M stars and brown dwarfs (termed ultracool dwarfs, UCDs) have been found to be the sources of unexpectedly intense radio emission at the frequencies of a few GHz (Berger \cite{ber05}; Hallinan et al. \cite{hal06}, \cite{hal07}, \cite{hal08}; Antonova et al. \cite{ant08}). The emission includes a slow-varying weakly-polarized component as well as short intense periodic bursts with almost 100\% polarization, whose period seems to coincide with the rotational period of the star. While the slow-varying component may be explained by the incoherent gyrosynchrotron radiation, the periodic bursts require a coherent radiation mechanism such as ECMI. Measurements of magnetic fields for the stars of late spectral class (M4) using phase-resolved spectropolarimetry (Donati et al. \cite{don06}, Morin et al. \cite{mor08}) indicate that the magnetic field has a dipole-like structure with the dipole axis close to the rotational axis. Thus the radio emission of UCDs seems to be similar to the auroral emissions of the Solar system planets, but with much a stronger magnetic field (not less than 3000 G, to provide the highest observed emission frequency). Existence of such magnetic fields at the cool stars ($<$M9) is also confirmed via infrared measurements (e.g., Reiners \& Basri \cite{rei07}).

ECMI has been investigated analytically and using numerical simulations in a number of papers (see, e.g., the review of Treumann \cite{tre06} and references therein). The required unstable electron distributions (with a deficiency of particles with low transversal velocity) are formed naturally due to particle reflection from a magnetic field gradient. Using a linear approximation, it is easy to show that even relatively weak fluxes of accelerated electrons can result in large growth rates, so the waves can be amplified from the level of thermal fluctuations up to the observed intensities at the distances not exceeding a few kilometers (see, e.g., Melrose \& Dulk \cite{mel82}, Bingham \& Cairns \cite{bin00}, Bingham, Cairns, \& Kellett \cite{bin01}, as well as estimations in this article). However, the resulting wave energy should become comparable with the particle energy; this results in a considerable particle diffusion on the waves, so the linear approximation becomes inapplicable. To interpret the cosmic radio emissions by ECMI, we have to use a nonlinear model that takes into account relaxation of unstable electron distribution due to interaction with the excited waves. In particular, only nonlinear models can provide us with such an important parameter as the transformation coefficient of the particle energy into waves.

Initially, ECMI was associated with the electron distribution of the loss-cone type. Kinetic simulation of relaxation of the loss-cone is made in works of Aschwanden (\cite{asc90}) and Fleishman \& Arzner (\cite{fle00}). However, in situ measurements in the AKR sources have shown that the radiation should be produced (at least, in this particular case) by ring-like or horseshoe-like distribution (e.g., Delory et al. \cite{del98}; Ergun et al. \cite{erg00}). These distributions are formed when electron beams (accelerated by an electric field) move into stronger magnetic field regions. Kinetic simulation of relaxation of such distributions have not been done yet (although there are some particle-in-cell simulations; see, e.g., the references in the review of Treumann \cite{tre06}).

When simulating relaxation of unstable electron distributions, the greatest challenge is spatial movement of the plasma waves and particles. Wave propagation out of the region occupied by the electrons with an unstable distribution naturally limits the wave amplification. On the other hand, the waves amplified in one part of the radiation source will cause relaxation of the electron distribution in other parts. Electron movement in an inhomogeneous magnetic field is necessary to form an unstable distribution (of the loss-cone or horseshoe type). To take into account all the mentioned factors, one has to create a 3D model of the emission source and adjacent regions (e.g., a magnetic loop in the solar or stellar corona) which, in turn, requires enormous computational resources. The most popular solution of this problem is to neglect the spatial movement of waves and particles completely, so that the model is reduced to an initial value problem where an arbitrary unstable electron distribution is used as the initial condition. We can neglect spatial movements if the time of wave/particle escaping from the wave amplification region $\tau_{\mathrm{esc}}$ far exceeds the typical diffusion time $\tau_{\mathrm{diff}}$. Such approach (a diffusive limit) is used in the above mentioned papers on kinetic simulation as well as in most particle-in-cell simulations. Obviously, this model is far from reality since it cannot explain long-term generation of emission and, in addition, requires almost instant formation of an unstable electron distribution. Nevertheless, a model considering only the wave growth/damping and particle diffusion allows us to (i) investigate the qualitative behaviour of relaxation of unstable electron distributions, (ii) determine the parameters of the produced waves, (iii) estimate the diffusion and relaxation timescales (which is necessary to make a conclusion about the validity of the diffusive limit), and (iv) determine the transformation coefficient of the accelerated particles energy into waves.

Measurements in the AKR sources have shown that the cold electron component is almost absent there. Under such conditions, the wave dispersion is determined mainly by the energetic electrons and the relativistic effects become important. Relativistic corrections to the dispersion relation result in a decrease in the cutoff frequency of the fast extraordinary mode; in addition, the fast and slow branches of the extraordinary mode can reconnect to form a single branch (Winglee \cite{win83}, \cite{win85}; Strangeway \cite{str85}, \cite{str86}; Robinson \cite{rob86}, \cite{rob87}; Le Qu\'eau \& Louarn \cite{leq89}; Louarn \& Le Qu\'eau \cite{lou96}). These effects allow the waves generated at the frequencies below the nonrelativistic cyclotron frequency to escape freely from the source into vacuum. Solving the exact relativistic dispersion equation is a complicated task. However, the existing studies (e.g., Robinson \cite{rob86}, \cite{rob87}) have shown that in a sufficiently hot low-density plasma, the wave dispersion becomes like that in vacuum (with the refraction index $N\to 1$ and group speed $\varv_{\mathrm{gr}}\to c$). Such an approximation seems to be valid in the sources of the planetary auroral radio emissions.

In this work, we have developed a kinetic relativistic 2D code for simulating the coevolution of electron distributions and plasma waves in the diffusive limit. Different wave modes can be considered both separately and simultaneously. Electron distribution of the horseshoe type is considered as the source of plasma oscillations. The calculations are made for the conditions that seem to be typical for UCDs, although the results can be easily scaled to other emission sources (e.g., the Earth or Jupiter). Exact plasma parameters in the magnetospheres of UCDs are unknown; in particular, we do not know whether the cold plasma component is present or not. Therefore we consider two opposite cases: (i) when a low-temperature plasma with the maxwellian distribution dominates and (ii) when such a component is absent. In the former case, we use the cold plasma dispersion relation; in the latter case, the dispersion relation is assumed to be like that in vacuum.

Note that the model used is restricted to the high-frequency elecromagnetic/magnetoionic waves and does not consider generation of the low-frequency waves (e.g., acoustic ones). In the AKR sources, amplitude of the low-frequency waves can reach high levels resulting in formation of solitary structures (such as electron and ion holes) which, in turn, can affect the generation of the radio emission (Pottelette, Treumann, \& Berthomier \cite{pot01}; Mutel et al. \cite{mut06}, \cite{mut07}). However, these effects are beyond the scope of this paper.

The model used is described in Section \ref{model}. The initial conditions of the model (including the electron distribution function) are described in Section \ref{initial}. The simulation results are presented in Section \ref{results} and discussed in Section \ref{discussion}. The conclusions are drawn in Section \ref{conclusion}. The calculation formulae (most of which can be found elsewhere) are given in Appendices.

\begin{table*}
\caption{Parameters of the different simulation models.}
\label{tab1}
\centerline{\begin{tabular}{cccccccccccc}
\hline
\rule{0pt}{10pt}Model & $f_{\mathrm{B}}$, Hz & $n_{\mathrm{b}}/n$ & $\omega_{\mathrm{p}}/\omega_{\mathrm{B}}$ & $E_{\mathrm{b}}$, keV & $\alpha_{\mathrm{c}}$ & $\gamma_{\max}$, $\textrm{s}^{-1}$ & $t_{\mathrm{ss}}$, s & $\tau_{\mathrm{sat}}$, s & $t_{\mathrm{ss}}\gamma_{\max}$ & $\tau_{\mathrm{sat}}\gamma_{\max}$ & $W_{\infty}/W_{\mathrm{b}0}$\\
\hline
\rule{0pt}{10pt}1  & $4\times 10^5$ & $10^{-2}$ & $10^{-2}$ & 10 & $60^{\circ}$  & $3.12\times 10^2$ & $2.09\times 10^{-1}$ & $1.97\times 10^{-1}$ & $65.2$ & $61.6$ & $0.118$\\
\rule{0pt}{10pt}2  & $4\times 10^5$ & 1         & $10^{-3}$ & 10 & $60^{\circ}$  & $3.48\times 10^2$ & $2.53\times 10^{-1}$ & $1.94\times 10^{-1}$ & $88.2$ & $67.4$ & $0.131$\\
\rule{0pt}{10pt}3  & $4\times 10^7$ & $10^{-2}$ & $10^{-2}$ & 10 & $60^{\circ}$  & $3.12\times 10^4$ & $1.82\times 10^{-3}$ & $1.73\times 10^{-3}$ & $56.8$ & $53.9$ & $0.119$ \\
\rule{0pt}{10pt}4  & $4\times 10^7$ & 1         & $10^{-3}$ & 10 & $60^{\circ}$  & $3.48\times 10^2$ & $1.65\times 10^{-3}$ & $1.64\times 10^{-3}$ & $57.3$ & $57.0$ & $0.132$\\
\rule{0pt}{10pt}5  & $4\times 10^9$ & $10^{-4}$ & $10^{-2}$ & 10 & $60^{\circ}$  & $3.12\times 10^4$ & $1.00\times 10^{-3}$ & $1.13\times 10^{-3}$ & $31.2$ & $35.1$ & $0.116$\\
\rule{0pt}{10pt}6  & $4\times 10^9$ & $10^{-4}$ & $10^{-1}$ & 10 & $60^{\circ}$  & $2.79\times 10^5$ & $1.51\times 10^{-4}$ & $2.79\times 10^{-4}$ & $42.2$ & $77.8$ & $0.107$\\
\rule{0pt}{10pt}7  & $4\times 10^9$ & $10^{-2}$ & $10^{-2}$ & 3  & $60^{\circ}$  & $5.33\times 10^6$ & $6.40\times 10^{-6}$ & $7.15\times 10^{-6}$ & $34.1$ & $38.1$ & $0.0841$\\
\rule{0pt}{10pt}8  & $4\times 10^9$ & $10^{-2}$ & $10^{-2}$ & 10 & 0             & $2.67\times 10^6$ & $1.80\times 10^{-5}$ & $2.09\times 10^{-5}$ & $48.0$ & $55.8$ & $0.103$\\
\rule{0pt}{10pt}9  & $4\times 10^9$ & $10^{-2}$ & $10^{-2}$ & 10 & $60^{\circ}$  & $3.12\times 10^6$ & $1.05\times 10^{-5}$ & $1.54\times 10^{-5}$ & $32.8$ & $48.2$ & $0.120$\\
\rule{0pt}{10pt}10 & $4\times 10^9$ & $10^{-2}$ & $10^{-2}$ & 10 & $90^{\circ}$  & $2.96\times 10^6$ & $2.21\times 10^{-5}$ & $1.99\times 10^{-5}$ & $65.5$ & $58.8$ & $0.115$\\
\rule{0pt}{10pt}11 & $4\times 10^9$ & $10^{-2}$ & $10^{-2}$ & 10 & $120^{\circ}$ & $1.88\times 10^6$ & $1.91\times 10^{-5}$ & $2.46\times 10^{-5}$ & $35.8$ & $46.2$ & $0.0552$\\
\rule{0pt}{10pt}12 & $4\times 10^9$ & $10^{-2}$ & $10^{-2}$ & 30 & $60^{\circ}$  & $1.15\times 10^6$ & $5.77\times 10^{-5}$ & $4.25\times 10^{-5}$ & $66.3$ & $48.8$ & $0.130$\\
\rule{0pt}{10pt}13 & $4\times 10^9$ & 1         & $10^{-3}$ & 3  & $60^{\circ}$  & $1.16\times 10^7$ & $3.12\times 10^{-6}$ & $4.39\times 10^{-6}$ & $36.2$ & $51.0$ & $0.133$\\
\rule{0pt}{10pt}14 & $4\times 10^9$ & 1         & $10^{-3}$ & 10 & 0             & $2.99\times 10^6$ & $2.19\times 10^{-5}$ & $2.24\times 10^{-5}$ & $65.5$ & $66.9$ & $0.115$\\
\rule{0pt}{10pt}15 & $4\times 10^9$ & 1         & $10^{-3}$ & 10 & $60^{\circ}$  & $3.48\times 10^6$ & $9.45\times 10^{-6}$ & $1.44\times 10^{-5}$ & $32.3$ & $50.2$ & $0.133$\\
\rule{0pt}{10pt}16 & $4\times 10^9$ & 1         & $10^{-3}$ & 10 & $90^{\circ}$  & $3.32\times 10^6$ & $9.82\times 10^{-6}$ & $1.74\times 10^{-5}$ & $32.6$ & $57.7$ & $0.129$\\
\rule{0pt}{10pt}17 & $4\times 10^9$ & 1         & $10^{-3}$ & 10 & $120^{\circ}$ & $2.21\times 10^6$ & $2.13\times 10^{-5}$ & $1.76\times 10^{-5}$ & $47.3$ & $39.0$ & $0.0621$\\
\rule{0pt}{10pt}18 & $4\times 10^9$ & 1         & $10^{-3}$ & 30 & $60^{\circ}$  & $1.17\times 10^6$ & $5.68\times 10^{-5}$ & $4.51\times 10^{-5}$ & $66.3$ & $52.7$ & $0.133$\\
\rule{0pt}{10pt}19 & $4\times 10^9$ & 1         & $10^{-2}$ & 10 & $60^{\circ}$  & $3.48\times 10^8$ & $1.65\times 10^{-7}$ & $1.64\times 10^{-7}$ & $57.3$ & $56.9$ & $0.132$\\
\hline
\end{tabular}}
\end{table*}

\section{Kinetic equations}\label{model}
If spatial movement of waves and particles is neglected, the coevolution of the electron distribution and waves in a plasma may be described in the general case by the following system of equations:
\begin{equation}\label{evolution}
\left\{\begin{array}{l}
\displaystyle\frac{\partial W^{(1)}_{\mathbf{k}}(\mathbf{k}, t)}{\partial t}=\gamma^{(1)}[\mathbf{k}, f(\mathbf{p}, t)]W^{(1)}_{\mathbf{k}}(\mathbf{k}, t),\\
\ldots\\
\displaystyle\frac{\partial W^{(N)}_{\mathbf{k}}(\mathbf{k}, t)}{\partial t}=\gamma^{(N)}[\mathbf{k}, f(\mathbf{p}, t)]W^{(N)}_{\mathbf{k}}(\mathbf{k}, t),\\
\displaystyle\frac{\partial f(\mathbf{p}, t)}{\partial t}=\frac{\partial}{\partial p_i}\left\{\sum\limits_{\sigma=1}^ND^{(\sigma)}_{ij}\left[\mathbf{p}, W^{(\sigma)}_{\mathbf{k}}(\mathbf{k}, t)\right]\frac{\partial f(\mathbf{p}, t)}{\partial p_j}\right\},
\end{array}\right.
\end{equation}
where $W_{\mathbf{k}}^{(\sigma)}$ is the energy density of oscillations of mode $\sigma$ in the space of wave vectors, $\sigma=1, \ldots, N$, $\gamma^{(\sigma)}$ is the growth rate for a given mode, $\mathbf{k}$ is the wave vector, $f$ is the electron distribution function, $D_{ij}^{(\sigma)}$ is the diffusion tensor (describing electron scattering on the waves of mode $\sigma$), and $\mathbf{p}$ is the electron momentum. The equations in the system (\ref{evolution}) are coupled implicitly, since the growth rate of plasma waves depends on the electron distribution function, while the diffusion tensor depends on the intensity and spectral distribution of the plasma waves. The expressions for the growth rate and elements of the diffusion tensor are given in Appendices \ref{increment} and \ref{diffusion}, respectively. 

In this work, we explore coevolution of the electron distribution and plasma waves using numerical simulations. The distributions $f(\mathbf{p})$ and $W_{\mathbf{k}}^{(\sigma)}(\mathbf{k})$ are defined on regular grids in $(p, \alpha)$- and $(\omega, \theta)$-spaces, respectively, where $\alpha$ is the electron pitch angle, $\omega$ is the wave frequency, and $\theta$ is the wave propagation direction (with respect to the ambient magnetic field). For different modes, different grids are used. In most simulations, we have considered two wave modes (e.g., ordinary + extraordinary). The grid size was chosen to be $60\times 60$ data points for all considered distributions. Note that in the work of  Aschwanden (\cite{asc90}), plasma waves were desribed using an irregular adaptive grid where the density of data points in phase space was approximately proportional to the initial growth rate. However, our simulation have shown that for the ring-like or horseshoe-like electron distributions, the region of positive growth in phase space can shift noticeably during the process of relaxation. Therefore, a regular grid is more suitable, and the considered area in $(\omega, \theta)$-space has to be wider than the initial region of positive growth. The system of equations (\ref{evolution}) is integrated with respect to time using the Gear formulae of fourth order (see Appendix \ref{numeric} for details of the numerical code). All cyclotron harmonics affecting the waves growth/damping and particle diffusion are considered; however, as a rule, the effect of the first harmonic is dominant.

Since the simulated system is closed, energy and particle number must be conserved in it (total system energy equals the sum of energy of particles and energies of all considered oscillation modes). Fulfillment of the conservation laws can be considered as a test of self-consistency in the model and accuracy of the numeric code. In our simulations, at the late stage of the relaxation process, the particle number was conserved with the relative error $\lesssim 1.5\times 10^{-3}$, and the total energy of the system was conserved with the relative error $\lesssim 3\times 10^{-3}$. These estimations correspond to the models without the thermal plasma component; for the models including the thermal plasma, the computation accuracy can be even better.

\section{Initial conditions}\label{initial}
In this work, we assume that the initial electron distribution function has the form:
\begin{equation}\label{f0b}
\left.f(\mathbf{p}, t)\right|_{t=0}=(n-n_{\mathrm{b}})f_0(\mathbf{p})+n_{\mathrm{b}}f_{\mathrm{b}}(\mathbf{p}),
\end{equation}
where $n$ is the total electron concentration, $n_{\mathrm{b}}$ is the concentration of nonthermal electrons, $f_0$ is the maxwellian distribution function of thermal electrons, and $f_{\mathrm{b}}$ is the distribution function of nonthermal electrons (both functions $f_0$ and $f_{\mathrm{b}}$ are assumed to be normalized to unity). Both the cases of $n_{\mathrm{b}}\ll n$ (thermal component dominates) and $n_{\mathrm{b}}=n$ (thermal component is absent) are considered.

\begin{figure*}
\centerline{\includegraphics{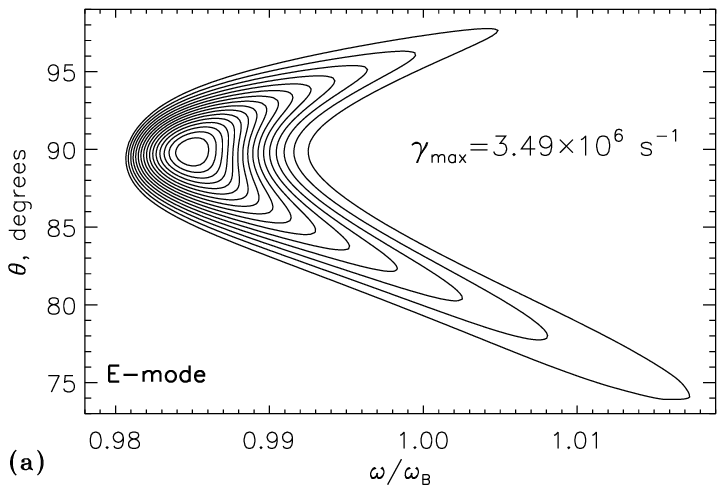}
\includegraphics{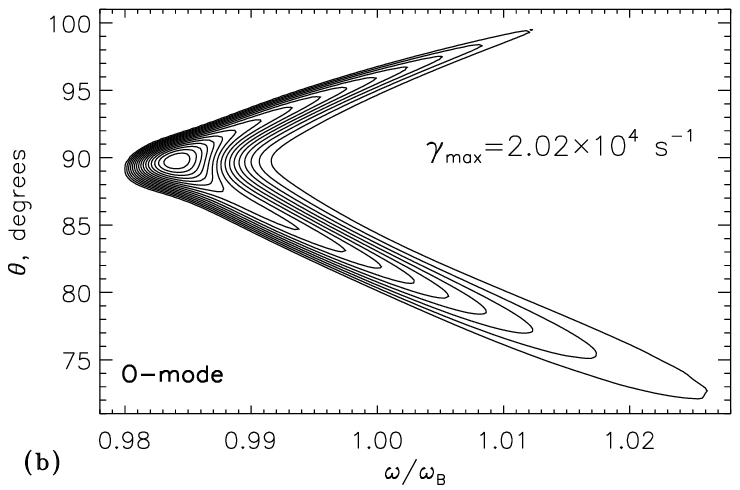}}
\caption{Initial growth rates of the extraordinary (a) and ordinary (b) waves. Simulation parameters correspond to the model 15 (Table \protect\ref{tab1}), and the wave dispersion is assumed to be like that in vacuum.}
\label{incr_V}
\end{figure*}

As stated above, we assume that the nonthermal electron distribution has the horseshoe-like shape (the distributions observed in the terrestrial magnetosphere are shown, e.g., at Figs. 3 and 5 in the paper of Ergun et al. \cite{erg00}). Instead of a detailed investigation of the formation process of a horseshoe-like distribution, we set the initial electron distribution to the model function similar to the observed ones:
\begin{equation}\label{fb}
f_{\mathrm{b}}(\mathbf{p})=A\exp\left[-\frac{(p-p_{\mathrm{b}})^2}{\Delta p_{\mathrm{b}}^2}\right]\left\{\begin{array}{ll}
1, & \mu\le\mu_{\mathrm{c}},\\[6pt]
\displaystyle\exp\left[-\frac{(\mu-\mu_{\mathrm{c}})^2}{\Delta\mu_{\mathrm{c}}^2}\right], & \mu>\mu_{\mathrm{c}},
\end{array}\right.
\end{equation}
where $A$ is the normalization factor and $\mu=\cos\alpha$. The shape of the distribution function is determined by such parameters as the typical electron momentum $p_{\mathrm{b}}$, electron dispersion in momentum $\Delta p_{\mathrm{b}}$, pitch-angle boundary of the loss-cone $\alpha_{\mathrm{c}}$ (or $\mu_{\mathrm{c}}=\cos\alpha_{\mathrm{c}}$), and the loss-cone boundary width $\Delta\mu_{\mathrm{c}}$. An example of the distribution function (\ref{fb}) can be seen at Fig. \ref{ev_b_V}a. For $\alpha_{\mathrm{c}}=0$, we obtain an isotropic ring-like distribution.

The initial energy density of plasma waves is assumed to equal the level of thermal oscillations:
\begin{equation}\label{W0}
\left.W_{\mathbf{k}}^{(\sigma)}(\mathbf{k}, t)\right|_{t=0}=\frac{k_{\mathrm{B}}T_0}{(2\pi)^3},
\end{equation}
where $k_{\mathrm{B}}$ is the Boltzmann constant and $T_0$ is the effective plasma temperature. In addition, it is assumed that the energy density of plasma waves cannot fall below the thermal level (\ref{W0}) during the process of wave/particle evolution, due to spontaneous radiation.

In all simulations, we assume that the nonthermal distribution function (\ref{fb}) has $\Delta p_{\mathrm{b}}/p_{\mathrm{b}}=0.2$ and $\Delta\mu_{\mathrm{c}}=0.2$. The thermal component of the plasma (if present) is described by a maxwellian distribution with temperature of $10^6$ K. The initial temperature of plasma waves $T_0$ equals $10^6$ K. The remaining parameters of the considered simulation models are given in Table \ref{tab1}; they were chosen in order to explore the influence of various factors on the process of wave/particle evolution. In all cases, the plasma density is relatively low, so that the ratio $\omega_{\mathrm{p}}/\omega_{\mathrm{B}}$ varies in the range from $10^{-3}$ to $10^{-1}$; the total electron concentration $n$ is calculated using the plasma frequency $\omega_{\mathrm{p}}$. The relative concentration of the energetic electrons $n_{\mathrm{b}}/n$ varies from $10^{-4}$ to 1.

\section{Results}\label{results}
\subsection{Relaxation of the horseshoe-like electron distribution}
\subsubsection{The case when a thermal plasma component is absent}\label{relaxation_V}
Firstly, we consider in detail an example of coevolution of the electron distribution and plasma waves. We assume here that a low-temperature thermal component is absent ($n_{\mathrm{b}}=n$). Investigations of the dispersion relations for weakly-relativistic plasmas (Robinson \cite{rob86}, \cite{rob87}) have shown that the wave dispersion (for the waves propagating across the magnetic field, $|\cos\theta|\ll 1$) becomes similar to that in vacuum if the typical electron speed $\varv_{\mathrm{e}}$ satisfies the condition $(\varv_{\mathrm{e}}/c)\gtrsim (\omega_{\mathrm{p}}/\omega_{\mathrm{B}})^2$. Such a requirement is satisfied, e.g., for the particle energy $E_{\mathrm{b}}\gtrsim 3$ keV and $\omega_{\mathrm{p}}/\omega_{\mathrm{B}}\lesssim 0.1$. Thus we assume that the wave refraction index equals unity both for the ordinary and extraordinary modes. Also we assume that the waves are elliptically polarized with the axial ratio of the polarization ellipse $T_{\mathrm{E}}=\cos\theta$ for the extraordinary mode and $T_{\mathrm{O}}=-1/\cos\theta$ for the ordinary mode ($T_{\mathrm{E}}T_{\mathrm{O}}=-1$). The above relations follow from the magnetoionic theory when $\omega_{\mathrm{B}}/\omega\to 1$ and $\omega_{\mathrm{p}}/\omega\to 0$ (Melrose \& Dulk \cite{mel91}; Willes, Melrose, \& Robinson \cite{wil94}); however, we found that the simulation results are not very sensitive to the exact value of the axial ratio $T_{\sigma}$ provided that for the quasi-transversal propagation $|T_{\mathrm{E}}|\ll 1$ (and, accordingly, $|T_{\mathrm{O}}|\gg 1$).

As an illustration, the following parameters were chosen: magnetic field $B=1430$ G that corresponds to the electron cyclotron frequency of $f_{\mathrm{B}}=4$ GHz (a typical value for the radio emission of ultracool dwarfs), plasma to cyclotron frequency ratio $\omega_{\mathrm{p}}/\omega_{\mathrm{B}}=10^{-3}$ that corresponds to the electron concentration $n=n_{\mathrm{b}}=2\times 10^5$ $\mathrm{cm}^{-3}$, typical energy of the energetic electrons $E_{\mathrm{b}}=10$ keV, and the loss-cone boundary $\alpha_{\mathrm{c}}=60^{\circ}$. In Table \ref{tab1}, this parameter set corresponds to the model 15.

Figure \ref{incr_V} shows the contours of growth rates of the ordinary and extraordinary modes at the initial moment ($t=0$). Only the positive growth rates are shown; the contour levels are evenly distributed between zero and the maximal growth rate value for a given mode which is also shown at the figure. One can see that the most effective wave amplification takes place slightly below the fundamental cyclotron frequency. Both emission modes are generated mainly in the perpendicular direction with respect to the magnetic field, with a slight asymmetry due to the loss-cone feature. Growth rates decrease rapidly with an increasing frequency, but the wave amplification (in an oblique direction) can occur even above the cyclotron frequency; in this region, the emission directivity patterns become essentially asymmetric. One can also note that the maximal growth rate of the extraordinary mode exceeds that of the ordinary mode by more than two orders of magnitude. Both for the ordinary and extraordinary modes, there are also amplification regions near higher harmonics of the cyclotron frequency, but with much lower growth rates (they are not shown at the figure). Thus the fundamental extraordinary mode is strongly dominating.

\begin{figure}
\centerline{\includegraphics{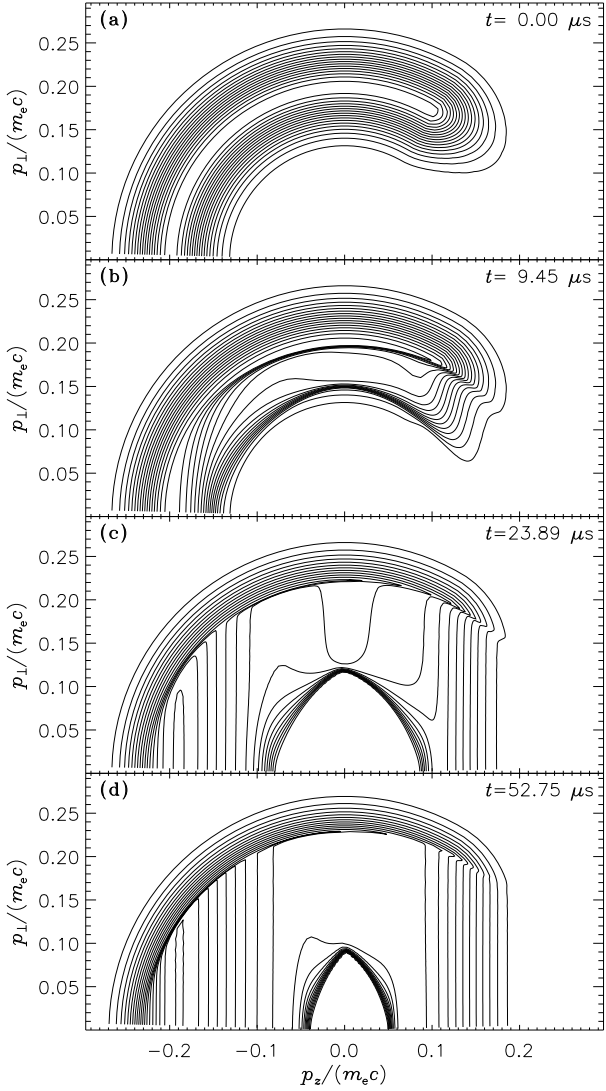}}
\caption{Time evolution of the electron distribution for the model 15 (Table \protect\ref{tab1}).}
\label{ev_b_V}
\end{figure}
\begin{figure}
\centerline{\includegraphics{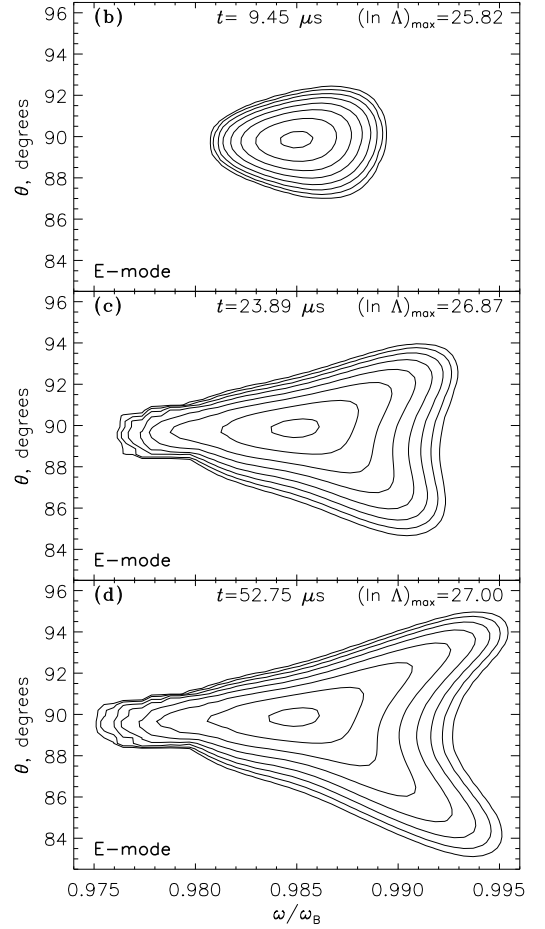}}
\caption{Time evolution of the extraordinary waves for the model 15 (Table \protect\ref{tab1}). The contour levels are 0.0075, 0.015, 0.03, 0.06, 0.125, 0.25, 0.5, and 0.9 of the maximum wave energy density.}
\label{ev_E_V}
\end{figure}

\begin{figure}
\centerline{\includegraphics{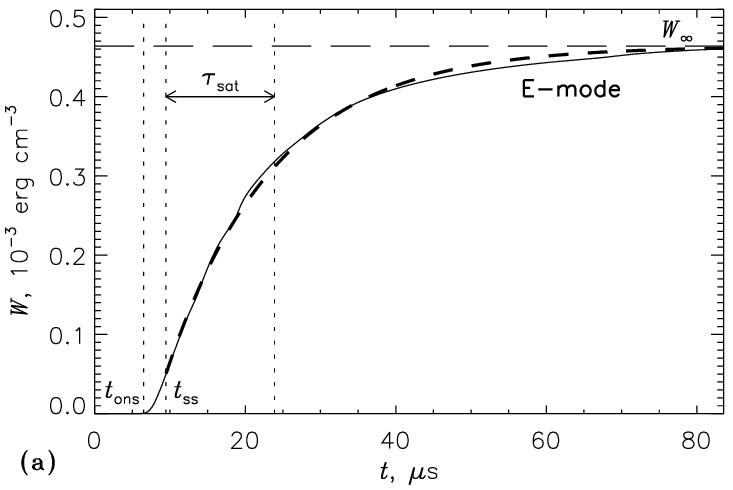}}
\centerline{\includegraphics{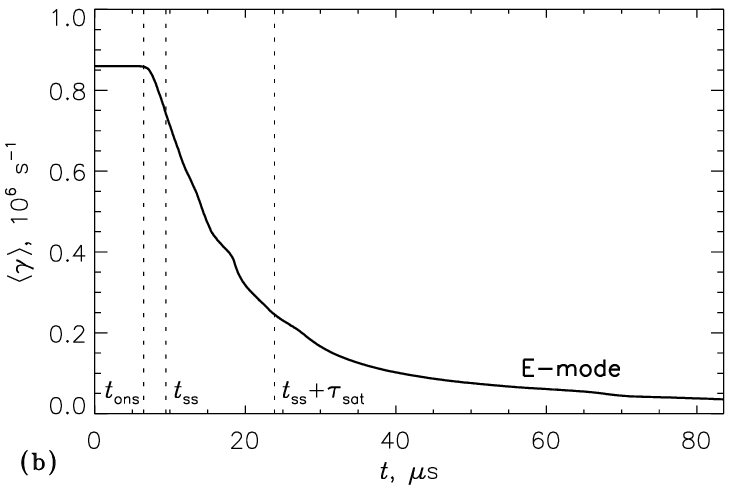}}
\caption{Time profiles of the total wave energy (a) and average growth rate (b). At panel (a), solid line is the result of numerical simulations and dashed line is a simplified functional fit. Simulation parameters correspond to the model 15 (Table \protect\ref{tab1}).}
\label{prof_V}
\end{figure}

Figure \ref{ev_b_V} shows the electron distribution function at different times (the chosen time moments correspond to the different stages of the relaxation process, see comment to Fig. \ref{prof_V}). For some time after the beginning of the simulation, while the wave energy density is low, the distribution function remains very similar to the initial one (Fig. \ref{ev_b_V}a). When the waves are amplified to a certain critical level, an effective diffusion of electrons on these waves begins. As a result, the electrons drift along the trajectories of $p_z=\textrm{const}$ towards lower values of $p_{\bot}$, thus losing energy. At the early relaxation stage (Fig. \ref{ev_b_V}b), this results in formation of a plateau slightly below the maximum of the initial distribution function (with respect to the momentum) and around $\alpha\simeq 90^{\circ}$ (with respect to the pitch angle). On the other hand, at the upper and lower boundaries of this plateau (with respect to the momentum), the gradient of the distribution function increases. At the middle relaxation stage (Fig. \ref{ev_b_V}c), the ``plateau'' expands with respect both to the momentum and pitch angle, and we can see that this feature is not really flat: while the derivative $\partial f/\partial p_{\bot}$ in this area approaches zero, the derivative $\partial f/\partial p_z$ remains nonzero, so the distribution function gradually decreases with increasing $p_z$. Nevertheless, since the maser amplification or damping of waves depends mainly on the derivative with respect to the transversal component of the momentum, a contribution of this region of the distribution function into growth rate is close to zero. At the upper boundary (with respect to the momentum), the ``plateau'' extends beyond the circle $p=p_{\mathrm{b}}$ (where the initial distribution function had a maximum) and the positive slope in $p_{\bot}$ disappears; at the lower boundary, this slope is still quite large. At the late relaxation stage (Fig. \ref{ev_b_V}d), the ``plateau'' expands even further, so now we have $\partial f/\partial p_{\bot}\le 0$ almost everywhere, and further amplification of waves nearly ceases. A positive slope remains only in the low-energy region, but the resulting growth rate of the waves is very low. Further simulation has not revealed any qualitatively new features: the ``hole'' around $p=0$ slowly shrinks and the distribution function approaches asymptotically a stationary (saturated) state, while its rate of change decreases with time.

Figure \ref{ev_E_V} shows the distribution of the extraordinary mode energy density in $(\omega, \theta)$-space at different times; in addition, the maximal values of the amplification coefficient $\Lambda$ are shown (this coefficient equals the ratio of the energy density at a given time to its initial level). The letters identifying the panels correspond to those in Fig. \ref{ev_b_V}, but panel (a) is omitted since the wave energy at the initial moment is negligible. For some time after the beginning of the simulation, the growth rate remains nearly constant, so the wave energy grows exponentially with time. As a result, at the onset of relaxation of the electron beam (Fig. \ref{ev_E_V}b), the waves are concentrated in a relatively narrow range with respect both to the frequency and the propagation direction; this region is much more narrow than the initial region of positive growth rate. When the logarithm of the amplification coefficient $\ln\Lambda$ reaches values of about 24-25, the waves begin to modify the electron distribution. This results in a sharp decrease (down to zero) of the growth rate in those regions where it was large at the beginning of the simulation, and also in an increase of the growth rate in the adjacent regions. These changes reflect formation of a plateau on the electron distribution function (see Fig. \ref{ev_b_V}b). As a result, further increase of the maximal intensity of waves nearly ceases, but the wave distribution in phase space broadens (see Fig. \ref{ev_E_V}c which corresponds to the middle relaxation stage). At a later stage (Fig. \ref{ev_E_V}d), the distribution of plasma waves becomes even more broad. Note that the shape of the final wave distribution differs considerably from the shape of the initial region of positive growth; in particular, relaxation of the electron beam allows waves to be generated at noticeably lower frequencies than at the initial moment. On the other hand, the wave energy density in the frequency range above the cyclotron frequency remains negligible throughout the relaxation process although the growth rate can be initially positive in this region.

Figure \ref{prof_V} shows the time history of the integral characteristics of the extraordinary waves: total energy density (integrated over the frequency and propagation angle) and the averaged growth rate (only the positive values of the growth rate were considered). In general, the time history is very similar to that obtained by Aschwanden (\cite{asc90}). For some time after the beginning of the simulation ($0<t\lesssim t_{\mathrm{ons}}$), the growth rate is constant and the wave energy grows exponentially but still remains very low. Then, when the wave energy reaches a certain critical level (at $t\simeq t_{\mathrm{ons}}$), relaxation of the unstable electron distribution begins and the growth rate starts to decrease. Note that this critical level is considerably lower than the final wave energy. After the onset of relaxation (at $t_{\mathrm{ons}}\lesssim t\le t_{\mathrm{ss}}$), the total energy of waves continues to grow with an increasing rate, but the corresponding curve somewhat differs from an exponential one. At time $t_{\mathrm{ss}}$ (which is defined as the time of the steepest slope), the rate of change of the total wave energy $\partial W/\partial t$ reaches its maximum. Later (at $t>t_{\mathrm{ss}}$), both the average growth rate and the rate of change of the total wave energy gradually decrease, approaching zero asymptotically; at the same time, the total wave energy goes asymptotically to the saturation level ($W_{\infty}$). At this stage, the time profile of the total wave energy can be described with a good accuracy by the well-known saturation curve:
\begin{equation}\label{saturation}
\displaystyle W(t)=W_{\mathrm{ss}}+(W_{\infty}-W_{\mathrm{ss}})\left[1-e^{-(t-t_{\mathrm{ss}})/\tau_{\mathrm{sat}}}\right],
\end{equation}
where $W_{\mathrm{ss}}=W(t_{\mathrm{ss}})$. The distributions of the electrons and waves at Figs. \ref{ev_b_V}-\ref{ev_E_V} correspond to the times $t=0$, $t=t_{\mathrm{ss}}$ (early stage of relaxation), $t=t_{\mathrm{ss}}+\tau_{\mathrm{sat}}$ (middle stage), and $t=t_{\mathrm{ss}}+3\tau_{\mathrm{sat}}$ (late stage). For the simulation parameters used in this section, relaxation begins at $t_{\mathrm{ons}}\simeq 6.5$ $\mu$s or $22.6\gamma_{\max}^{-1}$, where $\gamma_{\max}$ is the maximal growth rate of the extraordinary mode at the initial moment; the time of the steepest slope $t_{\mathrm{ss}}\simeq 9.45$ $\mu$s or $32.3\gamma_{\max}^{-1}$; and the relaxation timescale $\tau_{\mathrm{sat}}\simeq 14.4$ $\mu$s or $50.2\gamma_{\max}^{-1}$. The final (at $t\to\infty$) total energy density of the extraordinary waves equals $W_{\infty}\simeq 4.64\times 10^{-4}$ erg $\textrm{cm}^{-3}$, which amounts to 13.3\% of the energy density of the accelerated particles at the initial moment. The obtained timescales are similar (by order of magnitude) to those given in the article of Aschwanden (\cite{asc90}). However, the conversion efficiency of the particle energy into waves is now considerably higher, which is caused by the different type of unstable electron distribution (Aschwanden (\cite{asc90}) considered only the loss-cone).

As said before, growth rate of the ordinary mode is much lower than that of the extraordinary mode. Simulations considering both emission modes simultaneously have shown that the ordinary mode is amplified by less than a factor of 1.25 in comparison with the level of thermal oscillations. Thus the resulting energy of the ordinary waves is negligible and they neither make a measurable contribution to the radio emission of planets and UCDs nor affect the electron distribution.

\subsubsection{The case when a thermal plasma component is present}\label{relaxation_C}
Now we investigate the case when a thermal plasma component dominates. We use the same parameters of the magnetic field and energetic particles as in the previous section ($f_{\mathrm{B}}=4$ GHz, $n_{\mathrm{b}}=2\times 10^5$ $\textrm{cm}^{-3}$, $E_{\mathrm{b}}=10$ keV, $\alpha_{\mathrm{c}}=60^{\circ}$), but now a thermal plasma with the concentration of $n_0=2\times 10^7$ $\textrm{cm}^{-3}$ is present ($n_{\mathrm{b}}/n=10^{-2}$), so that the plasma to cyclotron frequency ratio is $\omega_{\mathrm{p}}/\omega_{\mathrm{B}}=10^{-2}$. In Table \ref{tab1}, this parameter set corresponds to the model 9. For the wave parameters, we use the cold plasma dispersion relation (see Appendix \ref{dispersion}).

\begin{figure*}
\centerline{\includegraphics{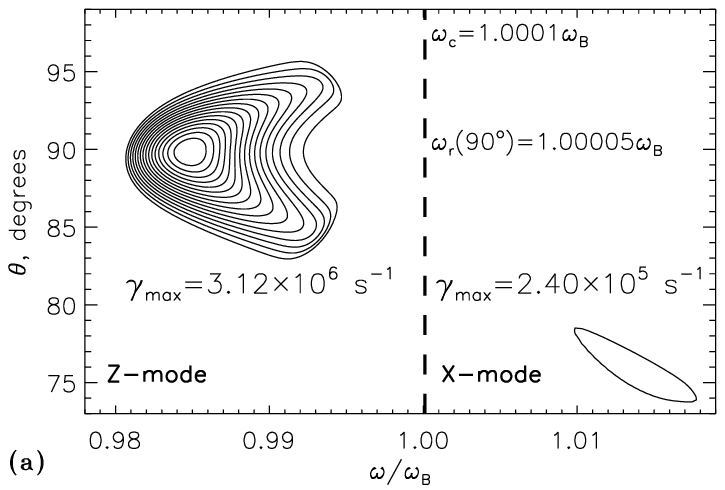}
\includegraphics{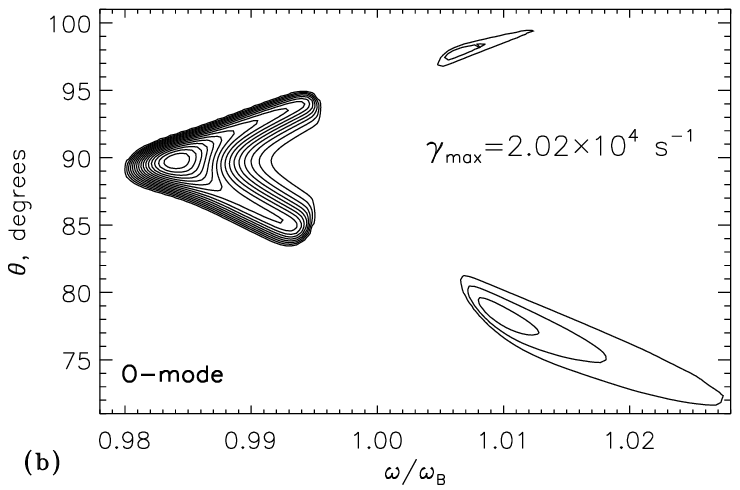}}
\caption{Initial growth rates of the extraordinary (a) and ordinary (b) waves in the presence of a thermal electron component. Simulation parameters correspond to the model 9 (Table \protect\ref{tab1}), and the cold plasma dispersion relation is used.}
\label{incr_C}
\end{figure*}

Figure \ref{incr_C} shows the initial growth rates (at $t=0$). In a cold magnetized plasma, the extraordinary mode is split into two branches: the fast (X) mode exists above the cutoff frequency $\omega_{\mathrm{c}}$, while the slow (Z) mode exists below the resonance frequency $\omega_{\mathrm{r}}$ ($\omega_{\mathrm{c}}>\omega_{\mathrm{r}}$). Under the conditions considered, both the cutoff and resonance frequencies almost coincide with the electron cyclotron frequency. However, the frequency gap between the amplification regions of the X- and Z-modes is much larger than $\omega_{\mathrm{c}}-\omega_{\mathrm{r}}$ since near the cyclotron frequency the waves are heavily damped by the thermal plasma component. The Z-mode is generated across the magnetic field, while the X-mode is generated in an oblique direction (at $\theta\simeq 75^{\circ}$). The maximal growth rate of the Z-mode exceeds that of the X-mode by more than an order of magnitude. The ordinary (O) mode has the smallest growth rate; its dispersion curve is continuous, but the amplification region is split into two by the absorption band near the cyclotron frequency.

\begin{figure}
\centerline{\includegraphics{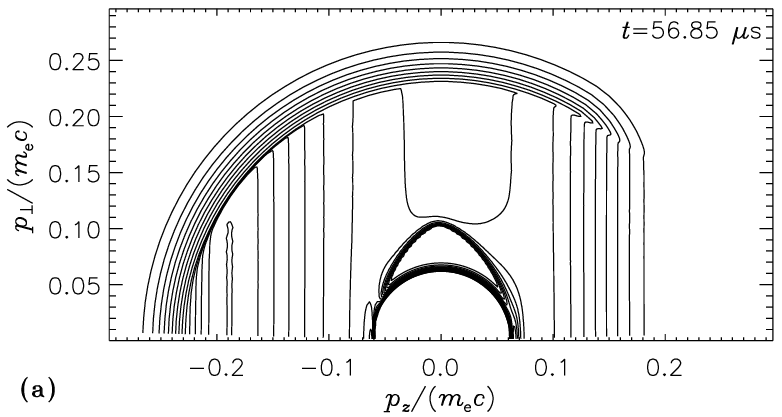}}
\centerline{\includegraphics{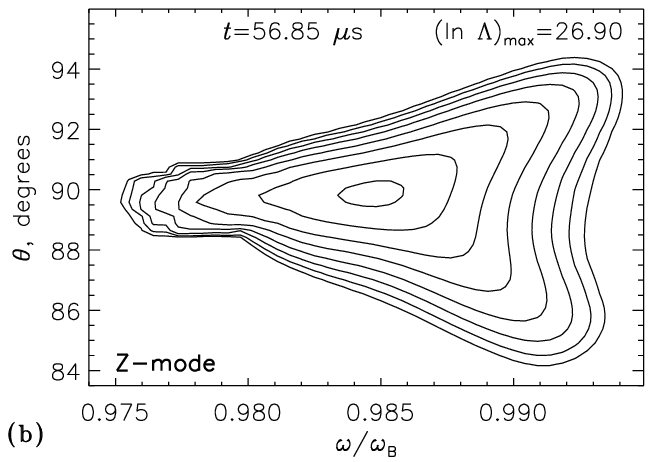}}
\caption{Electron distribution function (a) and the Z-mode energy density (b) at the late relaxation stage. The thermal component of the distribution function is shown only partially. Simulation parameters correspond to the model 9 (Table \protect\ref{tab1}).}
\label{fs_C}
\end{figure}

Simulations have shown that the time evolution of the electron distribution is very similar to that for the case without a thermal plasma (see Fig. \ref{ev_b_V}). Similarly, the time evolution of the Z-mode (which is the dominant mode) does not differ much from the evolution of the extraordinary mode shown in Fig. \ref{ev_E_V}. Therefore we do not present here a complete time history of the waves and particles. Figure \ref{fs_C} shows the late relaxation stage. One can see that the ``plateau'' on the electron distribution function (formed by the nonthermal electrons) merges with the thermal electron population at this stage. The conversion efficiency of the particle energy into the Z-mode waves (only the nonthermal component is considered) is about 11.8\%; this is slightly less that for the model without the thermal plasma. The relaxation timescales are very similar to the values found in the previous section. We can conclude that in a low-density plasma (with $\omega_{\mathrm{p}}/\omega_{\mathrm{B}}\ll 1$), the exact form of the dispersion relations is not very important, since both the vacuum-like and cold plasma dispersion relations provide almost the same results.

\begin{figure}
\centerline{\includegraphics{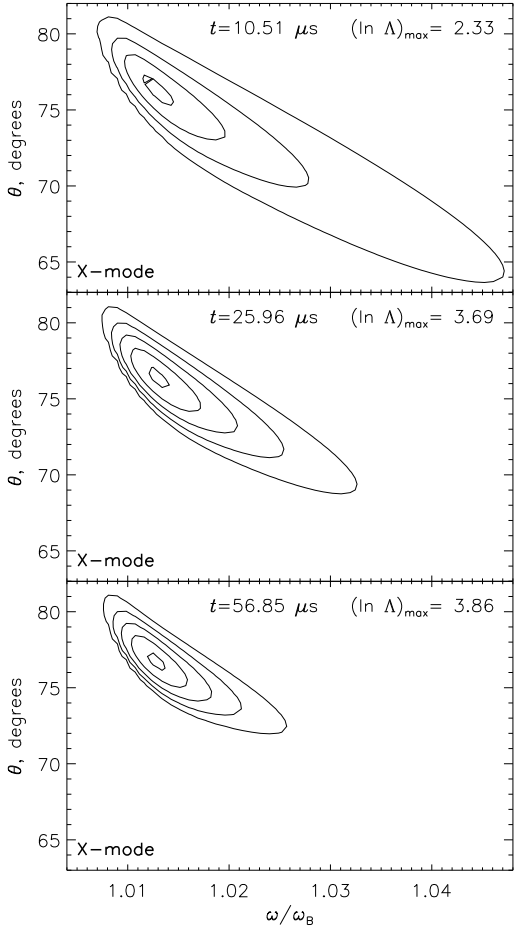}}
\caption{Time evolution of the X-mode waves for the model 9 (Table \protect\ref{tab1}). The contour levels are 0.125, 0.25, 0.5, and 0.9 of the maximum wave energy density.}
\label{ev_X_C}
\end{figure}

We can see that the generation of the Z-mode waves is very effective. However, these waves cannot escape from the generation region. Thus the question is of interest: can the intensities of the freely propagating modes (X- and/or O-mode) reach sufficiently high levels despite of a lower growth rate? We simulated the simultaneous evolution of the different modes. Figure \ref{ev_X_C} shows the distribution of the X-mode energy density at different times. The time history is different from that of the Z-mode since the X-mode generation is caused by the loss-cone feature. Since the loss-cone is destroyed during relaxation (due to diffusion of particles on the Z-mode waves), the X-mode amplification in some regions of phase space is replaced by absorption, so that the region occupied by the waves shrinks with time. Simulations for times later than those displayed in the figure have shown that the maximal intensity of waves also can somewhat decrease. However, the main result for the X-mode is that its intensity is much lower than that of the Z-mode: the parameter $\ln\Lambda$ does not exceed four, so the wave energy density exceeds the thermal level by not more than a factor of 50. Therefore the X-mode has no noticeable effect on relaxation of the electron distribution. 

\begin{figure}
\centerline{\includegraphics{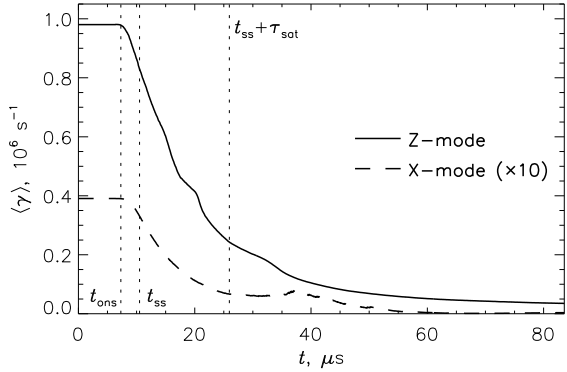}}
\caption{Time profiles of the average growth rates of the Z- and X-modes for the model 9 (Table \protect\ref{tab1}). For the X-mode, the scale is different from that for the Z-mode to make the growth rate variations more visible.}
\label{prof_C}
\end{figure}

Figure \ref{prof_C} shows the average growth rates of the Z- and X-modes. One can see that the growth rate of the X-mode starts to decrease simultaneously with that of the Z-mode. In general, the ratio of growth rates remains nearly the same throughout the relaxation process (despite a slight secondary increase of the average growth rate of the X-mode at the middle relaxation stage), so that the growth rate of the X-mode remains considerably lower than that of the Z-mode. We have found that the total energy density of the X-mode waves does not exceed $10^{-13}$ erg $\textrm{cm}^{-3}$, which amounts to $3\times 10^{-11}$ of the energy density of the accelerated particles. The energy density of the O-mode is much lower than that of the X-mode. 
Thus, in this simulation, relaxation of the unstable electron distribution is caused entirely by the Z-mode, and the intensities of the other wave modes are negligible.

\subsection{Comparison of the results for the different simulation models}
\begin{figure}
\centerline{\includegraphics{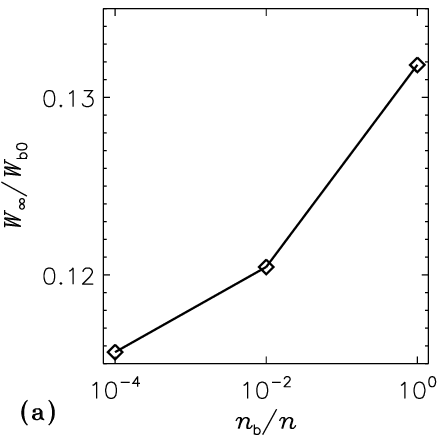}
\includegraphics{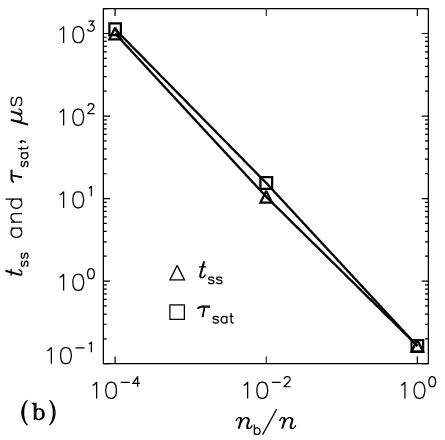}}
\caption{Energy conversion efficiency and the characteristic timescales of the relaxation process for the different concentrations of the energetic electrons. The values correspond to the models 5, 9, and 19 (Table \protect\ref{tab1}).}
\label{par_nb}
\end{figure}

Simulations of coevolution of the electron distributions and plasma waves were performed for the various parameter sets (see Table \ref{tab1}). The cold plasma and vacuum dispersion relations were used for the models with $n_{\mathrm{b}}/n\ll 1$ and $n_{\mathrm{b}}/n=1$, respectively. In all cases, the extraordinary mode (or its slow branch, for the models including a thermal plasma) dominated strongly. Table \ref{tab1} also contains the maximal initial growth rates, the timescales of the relaxation process, and the transformation coefficient of the particle energy into the waves. Figures \ref{par_nb}-\ref{par_fB} illustrate the effect of the various factors.

\subsubsection{Effect of varying the plasma parameters}
Figure \ref{par_nb} shows the simulation results (relaxation timescales and conversion efficiency of the particle energy into the waves) for the different relative concentrations of the energetic electrons ($n_{\mathrm{b}}/n=10^{-4}$, $10^{-2}$, and 1) while the total plasma density is assumed to be constant ($n=2\times 10^7$ $\textrm{cm}^{-3}$ that corresponds to $\omega_{\mathrm{p}}/\omega_{\mathrm{B}}=10^{-2}$). The typical energy of the accelerated electrons $E_{\mathrm{b}}$ and the loss-cone boundary  $\alpha_{\mathrm{c}}$ equal 10 keV and $60^{\circ}$, respectively. One can see that the relaxation timescales are simply inversely proportional to the concentration of the energetic particles. The conversion efficiency of the particle energy into the waves increases slightly with increasing $n_{\mathrm{b}}/n$ and varies from 11.2\% to 13.6\%.

\begin{figure}
\centerline{\includegraphics{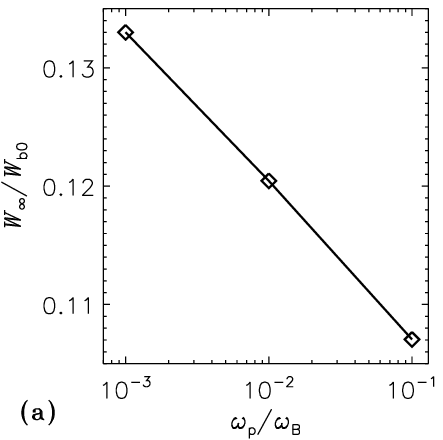}
\includegraphics{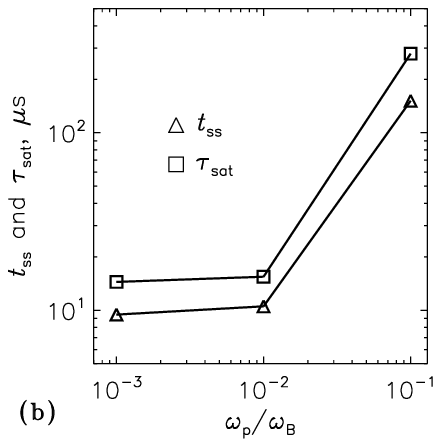}}
\caption{Same as in Fig. \protect\ref{par_nb}, for the different values of the total plasma density (concentration of the accelerated electrons $n_b$ is constant). The values correspond to the models 15, 9, and 6 (Table \protect\ref{tab1}).}
\label{par_Y}
\end{figure}

Figure \ref{par_Y} shows the simulation results for the case when the concentration of the energetic particles is constant ($n_{\mathrm{b}}=2\times 10^5$ $\textrm{cm}^{-3}$), while the total plasma density varies (which results in different ratios both of $n_{\mathrm{b}}/n$ and $\omega_{\mathrm{p}}/\omega_{\mathrm{B}}$). With increasing $\omega_{\mathrm{p}}/\omega_{\mathrm{B}}$ from $10^{-3}$ to $10^{-2}$, the growth rate of the extraordinary mode and the relaxation timescales remain nearly unchanged. At the same time, with increasing $\omega_{\mathrm{p}}/\omega_{\mathrm{B}}$ from $10^{-2}$ to $10^{-1}$, the growth rate decreases by about an order of magnitude (due to the changing dispersion parameters of the waves), while the relaxation timescales increase by the same factor. The conversion efficiency of particle energy into waves varies around 12\% and slightly decreases with increasing  $\omega_{\mathrm{p}}/\omega_{\mathrm{B}}$.

\subsubsection{Effect of varying the distribution of the energetic electrons}
\begin{figure}
\centerline{\includegraphics{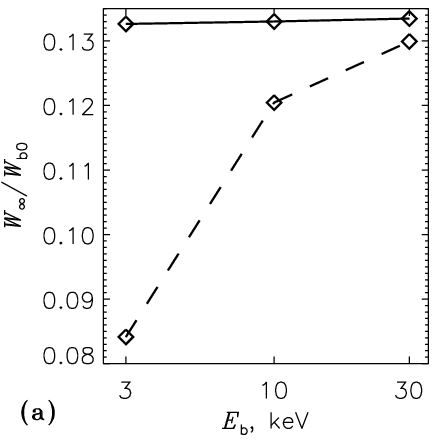}
\includegraphics{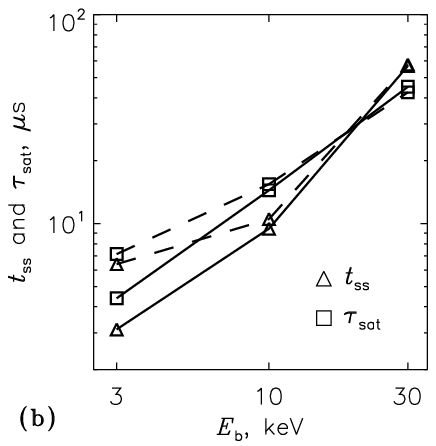}}
\caption{Same as in Fig. \protect\ref{par_nb}, for the different energies of the accelerated electrons. Solid lines: models 13, 15, and 18 (without thermal component); dashed lines: models 7, 9, and 12 (with thermal component).}
\label{par_Eb}
\end{figure}

\begin{figure}
\centerline{\includegraphics{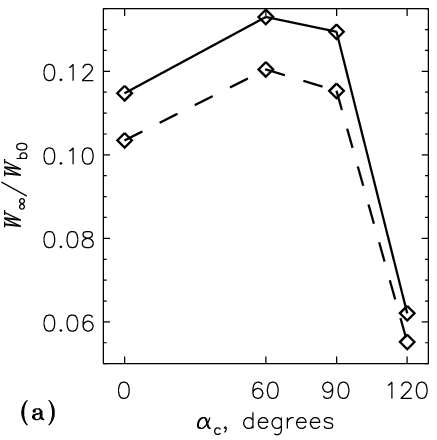}
\includegraphics{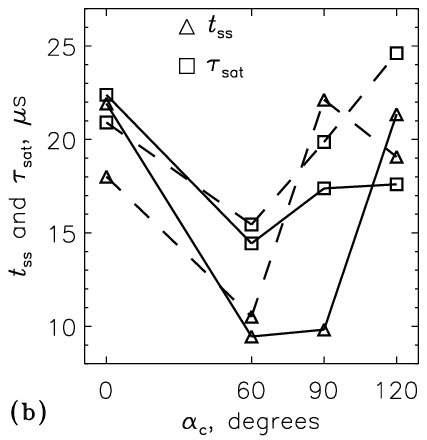}}
\caption{Same as in Fig. \protect\ref{par_nb}, for the different angular distributions of the accelerated electrons. Solid lines: models 14, 15, 16, and 17 (without thermal component); dashed lines: models 8, 9, 10, and 11 (with thermal component).}
\label{par_ac}
\end{figure}

In this section, we consider different distributions of the energetic electrons. In all cases, the concentration of the energetic electrons is assumed to be the same ($n_{\mathrm{b}}=2\times 10^5$ $\textrm{cm}^{-3}$). Both the models without a thermal plasma component and the models including such a component (with $n_{\mathrm{b}}/n=10^{-2}$) are considered; the corresponding points at Figs. \ref{par_Eb}-\ref{par_ac} are connected by the solid and dashed lines, respectively.

Figure \ref{par_Eb} shows the simulation results for the different typical energies of the accelerated particles (in all cases, the loss-cone boundary is $\alpha_{\mathrm{c}}=60^{\circ}$). For the models without a thermal plasma component, the conversion efficiency of the particle energy into the waves is almost constant (about 13.3\%). For the models including a thermal plasma,  the conversion efficiency increases with $E_{\mathrm{b}}$ (from 8\% at $E_{\mathrm{b}}=3$ keV to 13\% at $E_{\mathrm{b}}=30$ keV). An increase of the electron energy makes the relaxation process slower (both the timescales $t_{\mathrm{ss}}$ and $\tau_{\mathrm{sat}}$ increase). This is because the ``plateau'' formation for the electron distributions with higher energy requires a larger displacement of particles in the momentum space and, consequently, a higher energy density of plasma waves (in order to provide a stronger diffusion) and a longer time. For the models with a thermal component, the relaxation process is slower. With the increasing electron energy, the relaxation timescales (as well as the energy conversion efficiency) for the models with a thermal component approach the corresponding values for the models without a thermal component.

Figure \ref{par_ac} shows the simulation results for the different loss-cone boundaries (in all cases the beam energy is  $E_{\mathrm{b}}=10$ keV). The models without a thermal plasma component always provide a higher conversion efficiency of the particle energy into the waves than the corresponding models with a thermal component. The highest conversion efficiency as well as the fastest relaxation occur for $\alpha_{\mathrm{c}}=60^{\circ}$. For the ring-like distribution (with $\alpha_{\mathrm{c}}=0$), as well as for the loss-cone with $\alpha_{\mathrm{c}}=90^{\circ}$, the conversion efficiency and relaxation rate are slightly lower. The distribution with $\alpha_{\mathrm{c}}=120^{\circ}$ is similar to those used in the paper of Bingham \& Cairns (\cite{bin00}) and Bingham, Cairns, \& Kellett (\cite{bin01}); it has no electrons with pitch angles around $90^{\circ}$. Therefore, for this distribution, the amount of free energy is relatively low and only about 5-6\% of the initial electron energy can be transferred to waves. We would like to highlight that for all considered distributions (including essentially anisotropic ones), the waves are generated preferably in the perpendicular direction to the magnetic field; e.g., for the distribution with $\alpha_{\mathrm{c}}=120^{\circ}$, the maximum of the wave intensity (at the late relaxation stage) is at $\theta\simeq 91^{\circ}$.

\begin{figure}
\centerline{\includegraphics{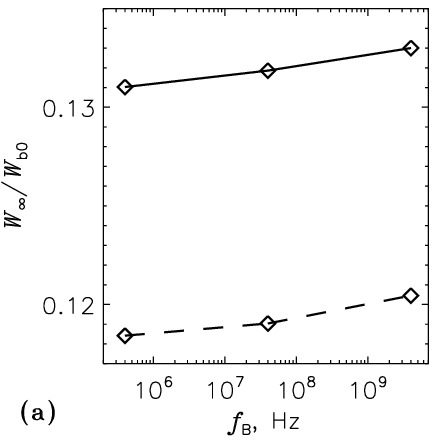}
\includegraphics{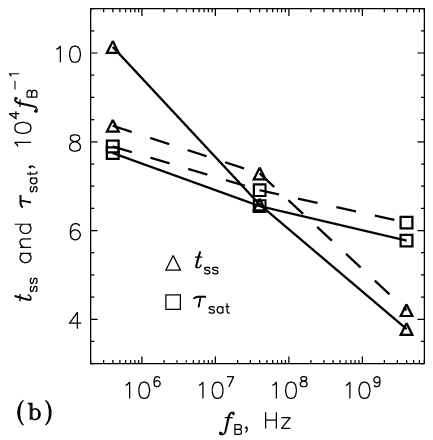}}
\caption{Same as in Fig. \protect\ref{par_nb}, for the different strengths of the magnetic field. Solid lines: models 2, 4, and 15 (without thermal component); dashed lines: models 1, 3, and 9 (with thermal component).}
\label{par_fB}
\end{figure}

\subsubsection{Effect of varying the magnetic field}
In this section, we investigate the influence of the magnetic field strength on the coevolution of the electron distributions and plasma waves. Figure \ref{par_fB} shows the simulation results for three values of the electron cyclotron frequency: 400 kHz (which is typical for the AKR sources in the terrestrial magnetosphere), 40 MHz (which is typical for the magnetosphere of Jupiter), and 4 GHz (which seems to be typical for the magnetospheres of UCDs). We assume that the plasma to cyclotron frequency ratio is constant and equals $\omega_{\mathrm{p}}/\omega_{\mathrm{B}}=10^{-3}$ for the models without a thermal plasma component; thus the concentrations of the energetic electrons $n_{\mathrm{b}}$ equal $2\times 10^{-3}$, $2\times 10^1$, and $2\times 10^5$ $\textrm{cm}^{-3}$, respectively. In the models including a thermal plasma, we assume that $\omega_{\mathrm{p}}/\omega_{\mathrm{B}}=10^{-2}$ and $n_{\mathrm{b}}/n=10^{-2}$ which provides the same concentrations of the energetic electrons as above. The typical energy of the accelerated electrons $E_{\mathrm{b}}$ and the loss-cone boundary  $\alpha_{\mathrm{c}}$ equal 10 keV and $60^{\circ}$, respectively. Under the conditions considered, the growth rates of the plasma waves are proportional to the cyclotron frequency (see Table \ref{tab1}). On the other hand, the relaxation timescales decrease with an increase of the cyclotron frequency somewhat faster than the inverse proportionality law implies. The largest relaxation timescales (relative to the inverse cyclotron frequency) occur at the Earth since in a weaker magnetic field diffusion of particles on plasma waves also weakens, and the relaxation onset requires a higher wave amplification coefficient $\Lambda$ (this effect is equivalent to that of reducing the initial temperature of plasma waves). The fastest relaxation (both in absolute and relative units) occurs for UCDs. A complete relaxation of the unstable electron distribution is achieved in less than 1 s (at the Earth), less than 10 ms (at Jupiter), and less than 0.1 ms (for UCDs). The conversion efficiency of the particle energy into the waves slightly increases with an increasing magnetic field strength; the models without a thermal plasma component always provide a higher conversion efficiency (13.1-13.3\%) than the corresponding models with a thermal component (11.8-12.0\%).

\section{Discussion}\label{discussion}
Our simulations have shown that in a relatively low-density plasma with  $\omega_{\mathrm{p}}/\omega_{\mathrm{B}}=10^{-3}-10^{-1}$, the ring-like or horseshoe-like distributions of accelerated electrons generate mainly the extraordinary waves with the frequency slightly below the electron cyclotron frequency. This conclusion is consistent with the results of previous studies (e.g., Ergun et al. \cite{erg00}). If the magnetoionic theory is applicable (i.e., the wave dispersion is dominated by a cold plasma component), the generated waves correspond to the Z-mode branch of the extraordinary mode. Generation of other modes (such as the ordinary mode, the X-mode of the magnetoionic theory, and the waves near higher cyclotron harmonics) is also possible, but their final energy density is less than that of the fundamental extraordinary mode by many orders of magnitude, even if the difference in the initial growth rates is not so large. We have found that the dominating mode remains the same throughout the relaxation process of an unstable electron distribution. This differs from the results of Fleishman \& Arzner (\cite{fle00}), where relaxation of a loss-cone distribution on the initially dominating mode (lower-hybrid waves) resulted in formation of a distribution which was stable with respect to excitation of lower-hybrid waves but still sufficiently anisotropic to amplify other types of waves (X and O). Most probably, this is because Fleishman \& Arzner (\cite{fle00}) considered a different electron distribution (loss-cone vs. horseshoe) and different plasma parameters  ($\omega_{\mathrm{p}}/\omega_{\mathrm{B}}>0.2$). In our simulations, relaxation of an electron distribution on the waves of the dominating mode prevented amplification of other modes as well. Also note that for other types of unstable electron distributions, another wave mode can prevail; for example, a loss-cone distribution in a low-density plasma excites mainly the oblique X-mode (Aschwanden \cite{asc90}).

For the ring-like or horseshoe-like electron distributions, the transformation coefficient of the particle energy into waves can exceed 10\%. This is considerably higher than for the loss-cone distributions (Aschwanden \cite{asc90}). The reason is that for the loss-cone, only a small fraction of particles (with pitch-angles around $\alpha_{\mathrm{c}}$) actually makes a contribution to the generation of waves and provides them with energy; in contrast, for the horseshoe-like distribution, the process of wave generation involves the majority of particles. It is interesting to note that the conversion efficiency is very weakly dependent on the plasma parameters and is determined mainly by the shape of the nonthermal distribution. The presence of a thermal plasma component reduces the conversion efficiency.

We estimate now the required efficiency of the emission mechanism. As an example, we consider the M9 dwarf TVLM 513--46546 located at a distance of $d=10.5$ pc. This object is known to produce radio bursts with an intensity up to $I=6$ mJy (at the frequency about 8 GHz) (Hallinan et al. \cite{hal07}). The typical size of the AKR generation region (in the direction across the magnetic field) can be estimated as 300 km. The radius of the UCD is expected to be comparable with that of Jupiter (i.e., ten times larger than the Earth), so we can assume that the radiation source has a size about $R_{\bot}\simeq 3000$ km. Thus the brightness temperature of the emission is about $10^{13}$ K. If the initial temperature of plasma waves equals $T_0=10^6$ K then the logarithm of the amplification coefficient should be not less than $\ln\Lambda\simeq 16$. For the growth rate  $\gamma\simeq 3\times 10^6$ $\textrm{s}^{-1}$ (see Section \ref{relaxation_V} for the beam and plasma parameters) and group velocity of the waves $\varv_{\mathrm{gr}}\le c$, such amplification can be achieved at a distance of no more than 1.6 km. Thus the estimated growth rates of the ECMI are well in excess of the required values, even for the relatively low concentrations of the accelerated electrons.

On the other hand, as stated above, estimations based only on the growth rate are insufficient and we need to check the conversion efficiency of the electron energy into radiation. The total power of radio emission from the UCD (for the case of isotropic radiation) can be estimated as $F_{\mathrm{r}}=4\pi d^2I\Delta f$, where $\Delta f$ is the spectral bandwidth of the emission. For the above parameters, assuming that $\Delta f\simeq f\simeq 8$ GHz (this is an upper limit as the spectrum is unlikely to be this broad), we obtain $F_{\mathrm{r}}\simeq 6\times 10^{21}$ erg $\textrm{s}^{-1}$. The energy flux of the accelerated electrons can be estimated as  $F_{\mathrm{b}}=n_{\mathrm{b}}\varv_{\mathrm{b}}E_{\mathrm{b}}R_{\bot}^2$, where $n_{\mathrm{b}}$, $\varv_{\mathrm{b}}$, and $E_{\mathrm{b}}$ are the concentration, speed, and energy of particles, and $R_{\bot}^2$ is the cross-section area of the electron beam. We assume that the plasma to cyclotron frequency ratio is $\omega_{\mathrm{p}}/\omega_{\mathrm{B}}=10^{-3}$ (which is less than that in the AKR sources) and all particles in the radiation source are accelerated ($n_{\mathrm{b}}=n$). Then for the cyclotron frequencies $f_{\mathrm{B}}=4-8$ GHz we obtain $n_{\mathrm{b}}\simeq (2-8)\times 10^5$ $\textrm{cm}^{-3}$. For the electron energy $E_{\mathrm{b}}=10$ keV, the corresponding speed $\varv_{\mathrm{b}}=5\times 10^9$ cm $\textrm{s}^{-1}$, and the beam size $R_{\bot}=3000$ km, the energy flux of the particles is  $F_{\mathrm{b}}\simeq (1.4-5.7)\times 10^{24}$ erg $\textrm{s}^{-1}$. Thus the conversion efficiency of the electron energy into the radio emission should be not less than $F_{\mathrm{r}}/F_{\mathrm{b}}\simeq (1-4)\times 10^{-3}$. According to our simulations, the ECMI efficiency ($\gtrsim 0.1$) is much larger. The above estimations do not take into account the possible absorption of the radio emission during propagation, radiation directivity, and (possibly) a narrow spectral band. Nevertheless, with reasonable assumptions about the source parameters, the ECMI is well able to provide the observed intensity of radio emission from UCDs. 

The terrestrial AKR is generated in the auroral cavity where a cold plasma component is almost absent; this allows the waves produced below the electron cyclotron frequency to escape directly from the source region due to relativistic corrections to the dispersion relation. It is very likely that the radio emission of UCDs is generated under similar conditions. However, let us discuss a possible case when a cold plasma component dominates. Under such conditions, the waves (of the Z-mode) excited by the ECMI cannot escape from the source due to a stop band at $\omega\simeq\omega_{\mathrm{B}}$. Thus the Z-mode has to be transformed into electromagnetic radiation, e.g., due to nonlinear processes. We can estimate the required efficiency of nonlinear conversion as $\eta_{\mathrm{NL}}\gtrsim (1-4)\times 10^{-2}$ (for the case when $\omega_{\mathrm{p}}/\omega_{\mathrm{B}}=10^{-2}$, $n_{\mathrm{b}}/n=10^{-2}$, 10\% of the particle energy goes into the Z-mode waves, and all other parameters are the same as in the previous paragraph). Nonlinear processes require a further investigation, but at present we cannot rule out the described two-stage emission mechanism. It is interesting to note that the growth rates of the X- and O-modes (see Fig. \ref{incr_C}), in a linear approximation, also can provide the required amplification coefficient at distances of tens to hundreds of kilometers. However, the kinetic simulations have shown that the conversion coefficient of the particle energy into these waves is extremely small, so a direct generation of electromagnetic waves by the horseshoe-like electron distribution in the presence of a cold plasma cannot be responsible for the observed emission.

According to the simulation results (see Table \ref{tab1}), the typical relaxation time of electron distributions with $\omega_{\mathrm{p}}/\omega_{\mathrm{B}}=10^{-2}$ and $n_{\mathrm{b}}/n\simeq 1$ in the terrestrial magnetosphere should be about $\tau_{\mathrm{diff}}\simeq\tau_{\mathrm{sat}}\simeq 2\times 10^{-3}$ s. On the other hand, the typical escape time of the waves from the generation region is $\tau_{\mathrm{esc}}\simeq R_{\bot}/\varv_{\mathrm{gr}}$. For $R_{\bot}\simeq 300$ km and $\varv_{\mathrm{gr}}\simeq c$, we obtain $\tau_{\mathrm{esc}}\simeq 10^{-3}$ s. Thus $\tau_{\mathrm{diff}}>\tau_{\mathrm{esc}}$, and the diffusive limit is, in fact, inapplicable. In the terrestrial magnetosphere, escape of waves from the amplification region has a dominating effect on the relaxation of the electron beams. In particular, this allows the strongly unstable electron distributions with large values of $\partial f/\partial p_{\bot}$ (such as reported by Ergun et al. \cite{erg00}) to exist in a quasi-stationary state. Escape of waves from the amplification region obviously reduces the relaxation rate, so that the processes of particle acceleration and magnetic mirroring are able to compensate the changes in the distribution function caused by diffusion on plasma waves. On the other hand, for UCDs we obtain $\tau_{\mathrm{esc}}\simeq 10^{-2}$ s, while $\tau_{\mathrm{diff}}\simeq 10^{-7}-10^{-5}$ s; thus $\tau_{\mathrm{diff}}\ll\tau_{\mathrm{esc}}$, and the use of a diffusive limit for UCDs is well justified. We can expect that the quasi-stationary electron distributions in the magnetospheres of UCDs will be very close to the relaxed state (as shown, e.g., at Figs. \ref{ev_b_V}c-d). To get more accurate results, one has to include the mechanism responsible for formation of an unstable electron distribution into the simulation model. The case when escape of waves from the amplification region is important for the relaxation process will be considered in a future paper.

Relaxation of unstable electron distributions in the magnetospheres of UCDs is very fast ($\ll 1$ s). On the other hand, the observed bursts of radio emission have a much longer duration (tens of seconds). It is very likely that the emission source is actually even more long-lived, but we observe the bursts only when the narrowly-directed radio beam (whose direction changes due to rotation of a star) points towards the observer (Hallinan et al. \cite{hal06}, \cite{hal07}; Berger et al. \cite{ber09}). This again means that the diffusive limit can be considered only as a very rough approximation. Actually, the emission properties are determined by the relatively stationary (but long-living) processes of particle acceleration; any model attempting to make a quantitative interpretation of observations has to include these processes. Nevertheless, we believe that the results obtained in this paper allow us to make some conclusions about the processes in magnetospheres of UCDs, and will be useful in developing more advanced models.

\section{Conclusion}\label{conclusion}
In this paper, we made a numerical simulation of the electron-cyclotron maser instability and investigated coevolution of unstable electron distribution and plasma waves in a low-density plasma ($\omega_{\mathrm{p}}/\omega_{\mathrm{B}}\ll 1$). Electron distribution of the ``horseshoe'' type was considered as an energy source for the plasma oscillations. The simulation parameters were chosen according to the measurements in the terrestrial magnetosphere, and scaled to the conditions expected in the magnetospheres of UCDs (where the electron cyclotron frequency is a few GHz). Spatial movement of the waves and particles, as well as other processes modifying the electron distribution (except of diffusion on plasma waves), were neglected. A number of simulations were made for the different parameters of plasma and accelerated particles. We found that:
\begin{itemize}
\item
Under the conditions considered, the electron beams with ring-like or horseshoe-like distributions generate mainly the extraordinary waves propagating across the magnetic field, with a frequency slightly below the electron cyclotron frequency. Other wave modes also can be amplified but their final intensity is less than that of the fundamental extraordinary mode by several orders of magnitude. 
\item
Conversion efficiency of the energy of accelerated electrons into the extraordinary waves can exceed 10\%; this parameter depends mainly on the shape of the nonthermal distribution. 
\item
A typical relaxation time of the unstable electron distribution is a few tens of inverse growth rates. In the magnetospheres of UCDs, a complete relaxation of the electron distribution should be very fast -- much less than one second; thus the observed emission requires a long-lived source of accelerated particles.
\item
Energy estimations show that the efficiency of the electron-cyclotron maser instability is sufficient to provide the observed intensity of radio emission from UCDs under reasonable assumptions about the energies and concentrations of accelerated particles.
\end{itemize}

\begin{acknowledgements}
The Armagh Observatory is supported by a grant from the Northern Ireland Dept. of Culture Arts and Leisure. We also thank the Leverhulme Trust for financial support, without whose funding this work would not have been possible.
\end{acknowledgements}

\appendix
\section{Dispersion of the magnetoionic modes}\label{dispersion}
The refraction index of the electromagnetic waves ($N_{\sigma}$) in a cold magnetized plasma satisfies the dispersion equation
\begin{equation}\label{N2}
N_{\sigma}^2=\left(\frac{kc}{\omega}\right)^2=1-\frac{2V(1-V)}{2(1-V)-U\sin^2\theta+\sigma\sqrt{{\mathcal D}}},
\end{equation}
where
\begin{equation}
{\mathcal D}=U^2\sin^4\theta+4U(1-V)^2\cos^2\theta,
\end{equation}
\begin{equation}
U=(\omega_{\mathrm{B}}/\omega)^2,\quad
V=(\omega_{\mathrm{p}}/\omega)^2,
\end{equation}
$\omega$ and $\mathbf{k}$ are the wave frequency and wave vector, $\omega_{\mathrm{p}}$ and $\omega_{\mathrm{B}}$ are the electron plasma and cyclotron frequencies respectively, and $\theta$ is the angle between the wave vector and the magnetic field. In Eq. (\ref{N2}), $\sigma=-1$ for the X-mode (fast extraordinary) and the Z-mode (slow extraordinary); $\sigma=+1$ for the O-mode (fast ordinary) and the W-mode (whistlers).

The polarization state of the magnetoionic modes is described by the following parameters:
\begin{equation}\label{Ts}
T_{\sigma}=\frac{2\sqrt{U}(1-V)\cos\theta}{U\sin^2\theta-\sigma\sqrt{\mathcal{D}}},
\end{equation}
\begin{equation}\label{Ls}
L_{\sigma}=\frac{V\sqrt{U}\sin\theta+T_{\sigma}UV\sin\theta\cos\theta}{1-U-V+UV\cos^2\theta},
\end{equation}
where $T_{\sigma}$ is the axial ratio of the polarization ellipse and $L_{\sigma}$ is the longitudinal part of the polarization.

The group velocity of the magnetoionic waves equals
\begin{equation}
\varv_{\mathrm{gr}}^{(\sigma)}=\frac{\partial\omega}{\partial k}=\frac{c}{\partial(\omega N_{\sigma})/\partial\omega},
\end{equation}
where
\begin{eqnarray}\label{N_dwN_dw}
N_{\sigma}\frac{\partial(\omega N_{\sigma})}{\partial\omega}&=&1+\frac{V\sqrt{U}T_{\sigma}\cos\theta}{2(T_{\sigma}-\sqrt{U}\cos\theta)^2}\nonumber\\
&\times&\left[1+\frac{(1+V)(1-T_{\sigma}^2)}{(1-V)(1+T_{\sigma}^2)}\right].
\end{eqnarray}
The derivative of the refraction index with respect to the propagation direction can be calculated using the relation
\begin{equation}\label{N_dN_dtheta}
\frac{1}{N_{\sigma}}\frac{\partial N_{\sigma}}{\partial\theta}=\frac{L_{\sigma}T_{\sigma}}{1+T_{\sigma}^2}.
\end{equation}

The different magnetoionic modes exist in the following frequency ranges:
\begin{itemize}
\item
X-mode: at $\omega>\omega_{\mathrm{c}+}$;
\item
O-mode: at $\omega>\omega_{\mathrm{p}}$;
\item
Z-mode: at $\omega_{\mathrm{c}-}<\omega<\omega_{\mathrm{r}+}$;
\item
whistlers: at $\omega<\omega_{\mathrm{r}-}$.
\end{itemize}
The cutoff and resonance frequencies are given by:
\begin{equation}
\omega_{\mathrm{c}\pm}=\pm\frac{1}{2}\omega_{\mathrm{B}}+\sqrt{\omega_{\mathrm{p}}^2+\frac{1}{4}\omega_{\mathrm{B}}^2},
\end{equation}
\begin{equation}
\omega_{\mathrm{r}\pm}^2=\frac{1}{2}(\omega_{\mathrm{B}}^2+\omega_{\mathrm{p}}^2)\pm\sqrt{\frac{1}{4}(\omega_{\mathrm{B}}^2+\omega_{\mathrm{p}}^2)^2-\omega_{\mathrm{B}}^2\omega_{\mathrm{p}}^2\cos^2\theta}.
\end{equation}

In vacuum, the dispersion parameters become:
\begin{equation}\label{NV}
N_{\sigma}=1,\quad
\varv_{\mathrm{gr}}^{(\sigma)}=c,\quad
\frac{\partial(\omega N_{\sigma})}{\partial\omega}=1,\quad
\frac{\partial N_{\sigma}}{\partial\theta}=0.
\end{equation}
By expanding the polarization parameters (\ref{Ts}-\ref{Ls}) in the small parameters $\sqrt{U}-1$ and $\sqrt{V}$ and retaining only the zero order terms in the expansions, we obtain the polarization state of the magnetioinic modes near the cyclotron frequency in a low-density plasma:
\begin{equation}\label{TV}
T_{-1}\simeq\cos\theta,\quad
T_{+1}\simeq -\frac{1}{\cos\theta},\quad
L_{\pm 1}\simeq 0.
\end{equation}

\section{Expressions for the growth rate}\label{increment}
According to Melrose \& Dulk (\cite{mel82}) and Aschwanden (\cite{asc90}), the growth rate of the magnetoionic oscillations equals (the wave-mode index $\sigma$ is hereafter omitted for brevity)
\begin{eqnarray}\label{incr1}
\gamma(\mathbf{k})&=&\frac{4\pi^2e^2c^2}{\omega N\partial(\omega N)/\partial\omega (1+T^2)}\nonumber\\
&\times&\sum\limits_{s=-\infty}^{\infty}\int\left[\frac{T(\cos\theta-N\beta_z)+L\sin\theta}{N_{\bot}\beta_{\bot}}J_s(\lambda)+
J'_s(\lambda)\right]^2\nonumber\\
&\times&\left[k_z\frac{\partial f(\mathbf{p})}{\partial p_z}+\frac{s\omega_{\mathrm{B}}}{\Gamma\varv_{\bot}}
\frac{\partial f(\mathbf{p})}{\partial p_{\bot}}\right]\beta_{\bot}^2\nonumber\\
&\times&\delta\left(\omega-k_z\varv_z-\frac{s\omega_{\mathrm{B}}}{\Gamma}
\right)\,\mathrm{d}^3\mathbf{p},
\end{eqnarray}
where $\mathbf{v}$ and $\mathbf{p}$ are the electron velocity and momentum, the indices $z$ and $\bot$ indicate the longitudinal and transversal components of the vectors with respect to the magnetic field (in particular, $N_z=N\cos\theta$ and $N_{\bot}=N\sin\theta$), $\beta=\varv/c$, $\Gamma=(1-\beta^2)^{-1/2}$ is the relativistic factor, $J_s(\lambda)$ and $J'_s(\lambda)$ are the Bessel function and its derivative over the argument $\lambda$,
\begin{equation}
\lambda=\frac{k_{\bot}p_{\bot}}{m_{\mathrm{e}}\omega_{\mathrm{B}}},
\end{equation}
and $f(\mathbf{p})$ is the electron distribution function satisfying the normalization condition
\begin{equation}
\int f(\mathbf{p})\,\mathrm{d}^3\mathbf{p}=n,
\end{equation}
where $n$ is the total electron concentration. The growth rates of the vacuum modes can be obtained by substituting the corresponding dispersion parameters (\ref{NV}-\ref{TV}).

If we introduce the dimensionless parameters
\begin{equation}
\mathbf{u}=\frac{\mathbf{p}}{m_{\mathrm{e}}c},\quad
x=\frac{\omega}{\omega_{\mathrm{B}}},\quad
Y=\frac{\omega_{\mathrm{p}}}{\omega_{\mathrm{B}}},
\end{equation}
and, in addition, use the polar coordinates $(u, \alpha)$ for the distribution function (where $\alpha$ is the angle between the electron velocity and the magnetic field), then the expression for the growth rate takes the form
\begin{eqnarray}\label{incr3}
\frac{\gamma(\mathbf{k})}{\omega_{\mathrm{B}}}&=&\frac{\pi Y^2}{N\partial(\omega N)/\partial\omega (1+T^2)}\nonumber\\
&\times&\sum\limits_{s=-\infty}^{\infty}\int
\left[\frac{T(\cos\theta-N\beta_z)+L\sin\theta}{N_{\bot}\beta_{\bot}}J_s(\lambda)+J'_s(\lambda)\right]^2\nonumber\\
&\times&\left[u_{\bot}\frac{\partial f(\mathbf{u})}{\partial u}+(\cos\alpha-N_z\beta)\frac{\partial f(\mathbf{u})}{\partial\alpha}\right]\frac{\sin\alpha}{\Gamma}\nonumber\\
&\times&\delta\left(x-\frac{s+xN_zu_z}{\Gamma}\right)\,\mathrm{d}^3\mathbf{u}.
\end{eqnarray}
The dimensionless distribution function $f(\mathbf{u})$ should satisfy the normalization condition
\begin{equation}
\int f(\mathbf{u})\,\mathrm{d}^3\mathbf{u}=1,
\end{equation}
and $\lambda=xN_{\bot}u_{\bot}$.

Using the properties of the $\delta$-function, we can reduce three-dimensional integrals in (\ref{incr3}) to one-dimensional integrals over $\mathrm{d}u_z$:
\begin{eqnarray}\label{incr4}
\frac{\gamma(\mathbf{k})}{\omega_{\mathrm{B}}}&=&\frac{2\pi^2Y^2}{xN\partial(\omega N)/\partial\omega (1+T^2)}\nonumber\\
&\times&\sum\limits_{s=-\infty}^{\infty}\int\limits_{u_{z\min}}^{u_{z\max}}
\left[\frac{T(\cos\theta-N\beta_z)+L\sin\theta}{N_{\bot}\beta_{\bot}}J_s(\lambda)+J'_s(\lambda)\right]^2\nonumber\\
&\times&\left[u_{\bot}\frac{\partial f(\mathbf{u})}{\partial u}+(\cos\alpha-N_z\beta)\frac{\partial f(\mathbf{u})}{\partial\alpha}\right]\nonumber\\
&\times&\left.\vphantom{\frac{1}{2}}\Gamma\sin\alpha\right|_{u_{\bot}=u_{\bot}(u_z)}\,\mathrm{d} u_z,
\end{eqnarray}
where the value $u_{\bot}$ at every point of the resonance curve is found from the resonance condition:
\begin{equation}\label{un_res}
u_{\bot}^2(u_z)=\left(\frac{s}{x}+N_zu_z\right)^2-u_z^2-1
\end{equation}
and the integration limits $u_{z\min}$ and $u_{z\max}$ (which are dependent on the harmonic number) are the boundaries of the interval where Eq. (\ref{un_res}) has a solution. 

For $|N_z|<1$ (which is always satisfied for X- and O-modes), the resonance curve in the momentum space (\ref{un_res}) is an ellipse intersecting the axis $u_{\bot}=0$ at two points:
\begin{equation}\label{uz12}
u_{z1,2}=\frac{sN_z/x\mp\sqrt{N_z^2+s^2/x^2-1}}{1-N_z^2},
\end{equation}
where $u_{z1}$ and $u_{z2}$ correspond to signs ``$-$'' and ``$+$'', respectively. In this case, $u_{z\min}=u_{z1}$ and $u_{z\max}=u_{z2}$. For $|N_z|>1$, Eq. (\ref{un_res}) describes a hyperbola with only one branch being physical (for which, at $u\to\infty$, the longitudinal component of the momentum $u_z$ has the same sign as $N_z$). Therefore the integration limits in (\ref{incr4}) will be equal (note that $u_{z1}>u_{z2}$ in this case): $u_{z\min}=-\infty$ and $u_{z\max}=u_{z2}$ at $N_z<0$; $u_{z\min}=u_{z1}$ and $u_{z\max}=\infty$ at $N_z>0$. Infinite integration limits can be avoided since the electron distribution function is usually defined only in a finite range of momentums, e.g., at $u\le u_{\mathrm{high}}$. In this case, the infinite limit (upper or lower, depending on the sign of $N_z$) should be replaced by the coordinate of the intersection point of the resonance curve with the circle $u=u_{\mathrm{high}}$, that is
\begin{equation}
u_{z\mathrm{high}}=\frac{\Gamma_{\mathrm{high}}-s/x}{N_z},
\end{equation}
where $\Gamma_{\mathrm{high}}$ is the relativistic factor of electrons with the momentum $u_{\mathrm{high}}$.

The above formulae for the growth rate involve infinite sums over cyclotron harmonics $s$. Actually, only the harmonics in a certain range $s_{\min}\le s\le s_{\max}$ make a contribution to the growth rate, since for the other harmonics either the resonance condition (for a given wave parameters) is never satisfied or the resonance curve lies outside the domain of the electron distribution function. In this work, the interval of harmonic numbers is taken with a large excess (say, from $-100$ to 100), and then we check for each harmonic whether it makes a contribution into the growth rate; this method is simple and very fast, and allows us to take into account all harmonics having an effect on the growth rate.

\section{Expressions for the quasi-linear diffusion rate}\label{diffusion}
The change in the electron distribution function due to diffusion on the magnetoionic waves is described by the last equation of the system (\ref{evolution}). In polar coordinates $(p, \alpha)$, this equation takes the form (Aschwanden \& Benz \cite{asc88})
\begin{eqnarray}\label{dif_eq1}
\frac{\partial f(\mathbf{p})}{\partial t}&=&\frac{1}{p^2\sin\alpha}\frac{\partial}{\partial\alpha}\left\{\sin\alpha\left[D_{\alpha\alpha}(\mathbf{p})\frac{\partial f(\mathbf{p})}{\partial\alpha}+pD_{\alpha p}(\mathbf{p})\frac{\partial f(\mathbf{p})}{\partial p}\right]\right\}\nonumber\\
&+&\frac{1}{p^2}\frac{\partial}{\partial p}\left\{p\left[D_{p\alpha}(\mathbf{p})\frac{\partial f(\mathbf{p})}{\partial\alpha}+pD_{pp}(\mathbf{p})\frac{\partial f(\mathbf{p})}{\partial p}\right]\right\},
\end{eqnarray}
where $D_{pp}$, $D_{p\alpha}$, $D_{\alpha p}$, and $D_{\alpha\alpha}$ are the components of the diffusion tensor ($D_{p\alpha}=D_{\alpha p}$). We note that the total diffusion tensor equals the sum of the diffusion tensors on separate modes (for simplicity, the formulae below refer only to one mode).

If we use the dimensionless momentum $\mathbf{u}$ and introduce, according to Aschwanden \& Benz (\cite{asc88}), generalized diffusion coefficients
\begin{equation}
D_0=\frac{D_{pp}}{(m_{\mathrm{e}}c)^2},\quad
D_1=\frac{D_{p\alpha}}{(m_{\mathrm{e}}c)^2}=\frac{D_{\alpha p}}{(m_{\mathrm{e}}c)^2},\quad
D_2=\frac{D_{\alpha\alpha}}{(m_{\mathrm{e}}c)^2},
\end{equation}
then Eq. (\ref{dif_eq1}) can be written in a form 
\begin{eqnarray}\label{dif_eq3}
\frac{\partial f(u, \alpha)}{\partial t}&=&\frac{1}{u^2}\left\{u\frac{\partial f(u, \alpha)}{\partial u}\right.\nonumber\\
&\times&\left.\left[2D_0(u, \alpha)+D_1(u, \alpha)\cot\alpha+\frac{\partial D_1(u,\alpha)}{\partial\alpha}\right]\right.\nonumber\\
&+&\left.\frac{\partial f(u, \alpha)}{\partial\alpha}\left[D_1(u, \alpha)+u\frac{\partial D_1(u, \alpha)}{\partial u}+D_2(u, \alpha)\cot\alpha\right]\right.\nonumber\\
&+&\left.2uD_1(u, \alpha)\frac{\partial^2f(u, \alpha)}{\partial u\partial\alpha}+u^2\frac{\partial}{\partial u}\left[D_0(u, \alpha)\frac{\partial f(u, \alpha)}{\partial u}\right]\right.\nonumber\\
&+&\left.\frac{\partial}{\partial\alpha}\left[D_2(u, \alpha)\frac{\partial f(u, \alpha)}{\partial\alpha}\right]\right\}.
\end{eqnarray}
Here the expression elements are rearranged to make the numerical calculations more convenient and accurate.

The diffusion coefficients $D_r$ (for an individual wave mode) are (Aschwanden \& Benz \cite{asc88}; Aschwanden \cite{asc90})
\begin{eqnarray}\label{dif1}
D_r(\mathbf{u})&=&\frac{4\pi^2e^2\sin^2\alpha}{m_{\mathrm{e}}^2c^2}\sum\limits_{s=-\infty}^{\infty}\int\left(\frac{\cos\alpha-N_z\beta}{\sin\alpha}\right)^r\nonumber\\
&\times&\frac{W_{\mathbf{k}}(\mathbf{k})}{N\partial(\omega N)/\partial\omega(1+T^2)}\nonumber\\
&\times&\left[\frac{T(\cos\theta-N\beta_z)+L\sin\theta}{N_{\bot}\beta_{\bot}}J_s(\lambda)+
J'_s(\lambda)\right]^2\nonumber\\
&\times&\delta\left(\omega-k_z\varv_z-\frac{s\omega_{\mathrm{B}}}{\Gamma}\right)\,\mathrm{d}^3\mathbf{k},
\end{eqnarray}
where $W_{\mathbf{k}}$ is the energy density of the considered mode in the space of wave vectors. The total energy density (the energy per unit volume) equals
\begin{equation}
W=\int W_{\mathbf{k}}(\mathbf{k})\,\mathrm{d}^3\mathbf{k}.
\end{equation}
The diffusion coefficients for the vacuum modes can be obtained by substituting the corresponding dispersion parameters (\ref{NV}-\ref{TV}).

If we use in (\ref{dif1}) the dimensionless frequency $x$ and the propagation direction $\theta$ as the wave characteristics instead of the wave vector, then the diffusion coefficients become 
\begin{eqnarray}\label{dif2}
D_r(\mathbf{u})&=&\omega_{\mathrm{B}}^2T_0\frac{k_{\mathrm{B}}e^2\sin^2\alpha}{m_{\mathrm{e}}^2c^5}\sum\limits_{s=-\infty}^{\infty}\int\left(\frac{\cos\alpha-N_z\beta}{\sin\alpha}\right)^r\nonumber\\
&\times&\left[\frac{T(\cos\theta-N\beta_z)+L\sin\theta}{N_{\bot}\beta_{\bot}}J_s(\lambda)+
J'_s(\lambda)\right]^2\nonumber\\
&\times&\frac{\mathcal{W}(x,\theta)Nx^2\sin\theta}{1+T^2}\delta\left(x-\frac{s+xN_zu_z}{\Gamma}\right)\,\mathrm{d}x\,\mathrm{d}\theta,
\end{eqnarray}
where $\mathcal{W}$ is the relative (with respect to the level of thermal oscillations) energy density of plasma waves in the space of wave vectors:
\begin{equation}
\mathcal{W}(x, \theta)=\frac{W_{\mathbf{k}}(x, \theta)}{W_{\mathbf{k}}^{(0)}},\quad
W_{\mathbf{k}}^{(0)}=\frac{k_{\mathrm{B}}T_0}{(2\pi)^3}.
\end{equation}
According to the initial conditions used, at the initial moment we have $\mathcal{W}(x, \theta)\equiv 1$.

The integral over $\mathrm{d}\theta$ in (\ref{dif2}) can be calculated analytically by using the properties of the $\delta$-function (see Aschwanden \& Benz \cite{asc88}; Aschwanden \cite{asc90}). As a result, the expression for the diffusion coefficients takes the form
\begin{eqnarray}\label{dif4}
\frac{D_r(\mathbf{u})}{\omega_B}&=&\omega_BT_0\frac{k_{\mathrm{B}}e^2}{m_{\mathrm{e}}^2c^5}\frac{\sin^2\alpha}{|\beta_z|}\sum\limits_{s=-\infty}^{\infty}\int\limits_{x_{\min}}^{x_{\max}}\left(\frac{\cos\alpha-N_z\beta}{\sin\alpha}\right)^r\nonumber\\
&\times&\left[\frac{T(\cos\theta-N\beta_z)+L\sin\theta}{N_{\bot}\beta_{\bot}}J_s(\lambda)+J'_s(\lambda)\right]^2\nonumber\\
&\times&\left.\frac{\mathcal{W}(x, \theta)x\sin\theta}{(1+T^2)|\sin\theta-(1/N)(\partial N/\partial\theta)\cos\theta|}\right|_{\theta=\theta(x)}\,\mathrm{d}x,
\end{eqnarray}
where the integration limits $x_{\min}$ and $x_{\max}$ correspond to the domain of the function $\mathcal{W}(x, \theta)$.

The resonance curve (dependence of $\theta$ on $x$) is found using the following considerations. If the particle momentum  $\mathbf{u}$ and wave frequency $x$ are known, then the resonance condition results in
\begin{equation}\label{N_z}
N_z=\frac{\Gamma-s/x}{u_z}.
\end{equation}
If the wave dispersion is assumed to be like in vacuum, then $N=1$ and we immediately obtain $\cos\theta=N_z$ (the requirement $|\cos\theta|\le 1$ must be satisfied). If the cold plasma dispersion relation is used, then the solution becomes a bit more complicated. With a known $N_z=N\cos\theta$, the dispersion equation (\ref{N2}) can be transformed to the biquadratic equation in the variable $\eta=\cos\theta$:
\begin{equation}\label{deq}
\mathcal{A}\eta^4+\mathcal{B}\eta^2+\mathcal{C}=0,
\end{equation}
where
\begin{equation}
\mathcal{A}=U-(1-V)^3-UV(1-N_z^2),
\end{equation}
\begin{equation}
\mathcal{B}=N_z^2[2(1-V)^2+UV(1-N_z^2)-2U],
\end{equation}
\begin{equation}
\mathcal{C}=N_z^4(U+V-1).
\end{equation}
This equation has a solution
\begin{equation}\label{dsol}
\eta^2=\frac{-\mathcal{B}\pm\sqrt{\mathcal{B}^2-4\mathcal{AC}}}{2\mathcal{A}}.
\end{equation}
In order to determine the correct sign (``$+$'' or ``$-$'') in the above formula for a given mode, one has to check whether the obtained values of the angle $\theta$ satisfy the dispersion equation (\ref{N2}). In practice, it is sufficient to check the validity of the inequality
\begin{equation}\label{modtest}
\sigma\left[\frac{2V(1-V)\eta^2}{\eta^2-N_z^2}-2(1-V)+U(1-\eta^2)\right]>0.
\end{equation}
In addition, the obvious requirements $\mathcal{B}^2-4\mathcal{AC}\ge 0$ and $\eta^2\le 1$ must be satisfied. Depending on the conditions, Eq. (\ref{deq}) for a given magnetoionic mode can have up to two solutions. If a solution with respect to  $\eta^2$ exists then the sign of $\eta$ coincides with that of $N_z$ (\ref{N_z}).

The interval of harmonic numbers contributing into the diffusion coefficients ($s_{\min}\le s\le s_{\max}$) can be estimated analytically, by using the resonance condition. As a result, for X- and O-modes (which have the refraction index $N\le 1$) we obtain
\begin{equation}
1\le s\le 2\Gamma x_{\max}.
\end{equation}
For the Z-mode and whistlers, the refraction index is not limited from above (in the cold plasma approximation), and we have to introduce such limits artificially: let $N\le N_{\max}$. Then we obtain
\begin{equation}
\Gamma x_{\max}(1-N_{\max})\le s\le \Gamma x_{\max}(1+N_{\max}).
\end{equation}
In this work, we use $N_{\max}=10$ which exceeds the actual values for the waves in the region of positive growth rate. The above formulae, as a rule, estimate the range of harmonic numbers with a large excess.

\section{Numeric code}\label{numeric}
In our simulations, the electron distribution function and energy density of plasma waves are represented as two-dimensional arrays:
\begin{equation}
\mathcal{W}(x, \theta)\to\mathcal{W}_{ij},\quad
f(u, \alpha)\to f_{mn},
\end{equation}
where $0\le i\le N_x-1$, $0\le j\le N_{\theta}-1$, $0\le m\le N_u-1$, and $0\le n\le N_{\alpha}-1$. As a result, the growth rate and diffusion coefficients need to be calculated only in the nodes of the corresponding grids, and these values (for each considered mode) can be represented as two-dimensional arrays as well:
\begin{equation}
\gamma(x, \theta)\to\gamma_{ij},\quad
D_r(u, \alpha)\to D^{(r)}_{mn}.
\end{equation}
Since the growth rate and diffusion coefficients are linearly dependent on the electron distribution function and wave energy density, respectively (see Eqs. (\ref{incr1}) and (\ref{dif1})), they can be computed using tensor multiplication (Fleishman \& Arzner \cite{fle00}): 
\begin{equation}\label{kerndef}
\gamma_{ij}=R_{ijmn}f_{mn},\quad
D^{(r)}_{mn}=P^{(r)}_{ijmn}\mathcal{W}_{ij},
\end{equation}
where the kernels $R_{ijmn}$ and $P^{(r)}_{ijmn}$ are constant throughout the simulation process. Therefore they can be computed once which reduces the computation time considerably. 

Since the resonance curves in the spaces of particle momentums and wave vectors, as a rule, do not pass through the grid points, the calculation of the growth rate and diffusion coefficients in the discretized case requires interpolating. In this work, we use a bilinear interpolation for the energy density of plasma waves. For the electron distribution function, we actually need to calculate its partial derivatives; this is done using a linear interpolation in one variable and a Lagrange interpolation on three closest points in another variable (where the derivative is needed).

For the electron distribution function, we use the grid where the momentum nodes are evenly distributed in the open interval from $u_{\mathrm{low}}$ to $u_{\mathrm{high}}$, and the pitch-angle nodes are evenly distributed in the open interval from 0 to $\pi$, so that
\begin{equation}
u_m=u_{\mathrm{low}}+\frac{u_{\mathrm{high}}-u_{\mathrm{low}}}{N_u}\left(m+\frac{1}{2}\right),\quad
\alpha_n=\frac{\pi}{N_{\alpha}}\left(n+\frac{1}{2}\right).
\end{equation}
For this grid, the best results are achieved (i.e., the conservation laws for the total energy and particle number are fulfilled with the highest accuracy) with the following boundary conditions (which are numerically implemented as an extrapolation of the arrays $f_{mn}$ and $D^{(r)}_{mn}$ beyond the grid):
\begin{equation}
\begin{array}{l}
\displaystyle f_{-1, n}=f_{0, n},\quad
f_{N_u, n}=f_{N_u-1, n},\\[5pt]
\displaystyle f_{m, -1}=f_{m, 0},\quad
f_{m, N_{\alpha}}=f_{m, N_{\alpha}-1},\\[5pt]
\displaystyle D^{(0)}_{-1, n}=D^{(0)}_{0, n},\quad
D^{(0)}_{N_u, n}=D^{(0)}_{N_u-1, n},\\[5pt]
\displaystyle D^{(1)}_{-1, n}=D^{(1)}_{0, n},\quad
D^{(1)}_{N_u, n}=D^{(1)}_{N_u-1, n},\\[5pt]
\displaystyle D^{(1)}_{m, -1}=-D^{(1)}_{m, 0},\quad
D^{(1)}_{m, N_{\alpha}}=-D^{(1)}_{m, N_{\alpha}-1},\\[5pt]
\displaystyle D^{(2)}_{m, -1}=-D^{(2)}_{m, 0},\quad
D^{(2)}_{m, N_{\alpha}}=-D^{(2)}_{m, N_{\alpha}-1}.
\end{array}
\end{equation}
For the energy density of plasma waves $\mathcal{W}_{ij}$, boundary conditions are not needed since the equations used do not contain derivatives of this value with respect to the wave parameters.
\end{document}